\documentclass[letterpaper]{article} % DO NOT CHANGE THIS
\usepackage{aaai2026}  % DO NOT CHANGE THIS
\usepackage{times}  % DO NOT CHANGE THIS
\usepackage{helvet}  % DO NOT CHANGE THIS
\usepackage{courier}  % DO NOT CHANGE THIS
\usepackage[hyphens]{url}  % DO NOT CHANGE THIS
\usepackage{graphicx} % DO NOT CHANGE THIS
\usepackage{natbib}  % DO NOT CHANGE THIS AND DO NOT ADD ANY OPTIONS TO IT
\usepackage{caption} % DO NOT CHANGE THIS AND DO NOT ADD ANY OPTIONS TO IT
\usepackage{algorithm}
\usepackage{algorithmic}

\usepackage{xcolor}
\newcommand{\answerYes}[1]{\textcolor{blue}{#1}} 
\newcommand{\answerNo}[1]{\textcolor{teal}{#1}} 
\newcommand{\answerNA}[1]{\textcolor{gray}{#1}}

\usepackage{newfloat}
\usepackage{listings}
\usepackage{subcaption}
\usepackage{multirow}
\usepackage{array}
\usepackage{enumitem}
\usepackage{xcolor}
\usepackage{tabularx}
\usepackage{amssymb}
\usepackage{amsmath}
\usepackage{booktabs}

\DeclareCaptionStyle{ruled}{labelfont=normalfont,labelsep=colon,strut=off} % DO NOT CHANGE THIS
\floatstyle{ruled}
\newfloat{listing}{tb}{lst}{}
\floatname{listing}{Listing}
\title{From Who They Are to How They Act: Behavioral Traits in Generative Agent-Based Models of Social Media}

\author{
Valerio La Gatta\textsuperscript{\rm 1},
Gian Marco Orlando\textsuperscript{\rm 2},
Marco Perillo\textsuperscript{\rm 2},
Ferdinando Tammaro\textsuperscript{\rm 2},
Vincenzo Moscato\textsuperscript{\rm 2}
}

\affiliations{
\textsuperscript{\rm 1}Northwestern University, Department of Computer Science, Evanston, IL, USA\\
valerio.lagatta@northwestern.edu
\\[0.5ex]
\textsuperscript{\rm 2}University of Naples Federico II, Department of Electrical Engineering and Information Technology, Naples, Italy\\
gianmarco.orlando@unina.it,
marco.perillo@unina.it,
fer.tammaro@studenti.unina.it,
vincenzo.moscato@unina.it
}

\begin{document}

\maketitle

\begin{abstract}
Generative Agent-Based Modeling (GABM) leverages Large Language Models to create autonomous agents that simulate human behavior in social media environments, demonstrating potential for modeling information propagation, influence processes, and network phenomena. While existing frameworks characterize agents through demographic attributes, personality traits, and interests, they lack mechanisms to encode behavioral dispositions toward platform actions, causing agents to exhibit homogeneous engagement patterns rather than the differentiated participation styles observed on real platforms. In this paper, we investigate the role of behavioral traits as an explicit characterization layer to regulate agents' propensities across posting, re-sharing, commenting, reacting, and inactivity. Through large-scale simulations involving 980 agents and validation against real-world social media data, we demonstrate that behavioral traits are essential to sustain heterogeneous, profile-consistent participation patterns and enable realistic content propagation dynamics through the interplay of amplification- and interaction-oriented profiles. Our findings establish that modeling \textit{how agents act}—not only \textit{who they are}—is necessary for advancing GABM as a tool for studying social media phenomena.
\end{abstract}

% Uncomment the following to link to your code, datasets, an extended version or similar.
% You must keep this block between (not within) the abstract and the main body of the paper.
% \begin{links}
%     \link{Code}{https://aaai.org/example/code}
%     \link{Datasets}{https://aaai.org/example/datasets}
%     \link{Extended version}{https://aaai.org/example/extended-version}
% \end{links}

\section{Introduction}

Generative Agent-Based Modeling (GABM) has emerged as a powerful paradigm for simulating social systems by leveraging Large Language Models (LLMs) to create autonomous agents that exhibit human-like behavior without rigid, predefined rules \cite{park2023generative}. Unlike traditional agent-based models that rely on hand-crafted behavioral rules, GABM agents can reason about their environment, plan actions, and interact flexibly based on their characterization and experiences, enabling more naturalistic simulations of complex social dynamics \cite{naveed2025comprehensive}.

In the context of social media, GABM has shown particular promise for reproducing online dynamics, including information propagation \cite{zhang2025ga}, influence processes \cite{touzelsimulation}, and the emergence of polarization \cite{ohagi-2024-polarization} and echo chambers \cite{ferraro2024agent, wang2025decoding}. Moreover, GABM has proven capable of replicating well-known network-level phenomena, such as the friendship paradox \cite{orlando2026validating}.

{\color{black}Despite these promising results, the value of GABM as a tool for studying web and social media depends on agents reproducing the behavioral heterogeneity of real user populations; a condition that existing applications struggle to meet, as they remain limited in how agents are characterized.} Agents are typically initialized with identity profiles derived either from real-world datasets \cite{tornberg2023simulating} or synthetically generated \cite{li2023you}, encoding biographical attributes such as age, gender, ethnicity, education, or occupation. While such information establishes \textit{who agents are}, it does not specify \textit{how they act}—that is, their behavioral dispositions toward different types of platform actions. As a consequence, agents with diverse identities often exhibit homogeneous engagement patterns, predominantly converging toward content generation while neglecting other participation modes such as re-sharing, commenting, or selective inactivity \cite{ferraro2024agent}. This behavioral homogeneity limits current approaches' ability to reproduce the differentiated participation styles—lurkers, amplifiers, contributors, engagers—that characterize real social platforms. To our knowledge, the only work that explicitly introduces behavior-oriented profiles in LLM-based simulations is \cite{wang2025user}, which defines prototypical user roles tailored to recommender systems. While this represents an important step toward diversifying agent behaviors, its focus remains on role-driven content consumption within the recommendation domain, rather than regulating participation patterns across the full action space of social media platforms.

Our work addresses this gap by formalizing \textit{behavioral traits} as an explicit characterization layer that regulates agents' propensities across the complete social media action space: posting original content, re-sharing others' posts, commenting, reacting (liking/disliking), and remaining inactive. Grounded in empirical user typologies from online participation research \cite{akar2018user, lcp_segmentation, temporal_posting_behavior, murdock2024agent, mut_typology}, our seven archetypal behavioral profiles—\textit{Silent Observers}, \textit{Occasional Sharers}, \textit{Occasional Engagers}, \textit{Balanced Participants}, \textit{Content Amplifiers}, \textit{Proactive Contributors}, and \textit{Interactive Enthusiasts}—instantiate distinct participation patterns that differ in activity frequency and interaction style.

We investigate whether explicit behavioral traits are necessary for generative agents to sustain heterogeneous participation patterns, reproduce emergent propagation dynamics, and align with structural roles observed in real social media platforms. We operationalize these objectives through four research questions:

\begin{itemize}
    \item[\textbf{RQ1:}] \emph{How do behavioral traits influence the emergence of heterogeneous agent behaviors?}
    \item[\textbf{RQ2:}] \emph{How do behavioral traits affect content propagation?}
    \item[\textbf{RQ3:}] \emph{How do behavioral traits affect agents' network centrality?}
    \item[\textbf{RQ4:}] \emph{How well do behavioral traits reproduce real-world social networks?}
\end{itemize}

We extend an existing GABM framework \cite{orlando2026validating} to incorporate the above-mentioned behavioral traits alongside identity profiles. {\color{black}Through large-scale simulations involving 980 agents, we investigate whether identity-only and personality descriptors from the OCEAN model can induce behavioral differentiation in generative agents.

Our results demonstrate that behavioral traits are essential for reproducing realistic social media dynamics:} \textit{(i)} they prevent behavioral homogeneity by sustaining heterogeneous, profile-consistent participation patterns; \textit{(ii)} enable content propagation cascades through the interplay of amplification-oriented profiles (Content Amplifiers and Occasional Sharers); \textit{(iii)} shape distinct structural roles in emergent networks, with amplification-oriented agents dominating re-sharing networks and interaction-oriented agents (Interactive Enthusiasts, Occasional Engagers) leading interaction networks; and \textit{(iv)} successfully reproduce network structures observed in real-world social media when grounded in empirical data. These findings establish that modeling \textit{how agents act}—not only \textit{who they are}—is essential for advancing GABM as a tool for studying social media phenomena. {\color{black}To ensure transparency and reproducibility, we publicly release our code\footnote{{\color{black}\url{https://github.com/PRAISELab-PicusLab/LLM-Agents-withBehavioralTraits-Simulation-Framework}}}.}

\section{Related Work}\label{sec:related}

\subsection{Generative Agent-Based Modeling}

LLMs have recently demonstrated exceptional capabilities in emulating human-like decision-making and social behavior, opening new avenues for agent-based simulations. \citet{park2023generative} pioneered Generative Agent-Based Modeling (GABM) by proposing an architecture for \textit{generative agents}, LLM-driven autonomous entities that can store and recall experiences, reflect on them, and plan future actions within a simulated environment. Given these capabilities, GABM offer a powerful way to model online social phenomena such as social influence \cite{touzelsimulation}, information diffusion \cite{tornberg2023simulating}, and opinion formation \cite{ashery2025emergent}. {\color{black}Recent efforts have produced increasingly sophisticated simulation infrastructures: \citet{rossetti2024ysocial} introduces a comprehensive social media digital twin featuring a rich action space, multiple recommender systems, and agent characterization grounded in demographic attributes and OCEAN personality traits \cite{mccrae1992introduction}.

Despite these advancements, most existing GABM simulations define agents solely through personality descriptions \cite{rossetti2024ysocial, huang2026designing}, causing agents with distinct identities to still act alike and narrowing both the range of emergent dynamics the simulation can produce and the phenomena it can meaningfully study.} Since network-level properties such as content diffusion and structural centrality emerge from heterogeneous individual behaviors \cite{jackson2013diffusion, zhou2024beyond}, the absence of differentiated behavioral dispositions restricts the analytical scope of prior work.
\emph{Our study addresses this gap by introducing \textit{behavioral traits} as an explicit dimension of agent characterization that regulates participation patterns across the platform's action space.}

\subsection{Modeling Online Human Behavior}

Research on online participation has long emphasized the heterogeneity of user engagement patterns, identifying diverse roles such as lurkers, amplifiers, contributors, and debaters \cite{akar2018user}. This behavioral diversity fundamentally shapes information diffusion, engagement dynamics, and network structure in online communities. Early agent-based models captured such heterogeneity by hard-coding or statistically calibrating agent types based on empirical user typologies \cite{murdock2024agent}, combining explicit attributes (demographics) with implicit behavioral cues derived from digital traces like posting frequency and interaction patterns \cite{wang2021user}. The advent of LLMs has enabled more flexible modeling approaches \cite{takata2024spontaneous, newsham2025personality}, giving rise to Generative Agent-Based Modeling (GABM) for social media simulation.

{\color{black}However, existing GABM approaches typically rely on identity profiles—demographic attributes, personality descriptions, topical interests, or role-based personas—as the primary characterization dimension, leaving behavioral dispositions underspecified \cite{ferraro2024agent, orlando2026validating, ohagi-2024-polarization, zhang2025ga, coop_behavior, ji2025llm, farr2025simulating}.} Consequently, agents with distinct identities often exhibit homogeneous engagement patterns, predominantly converging toward content generation while neglecting other participation modes. {\color{black}Recent hybrid solutions \cite{mou2024unveiling, ng2025llm} address this by assigning predefined communication roles to influential users or fixing per-agent activity counts. Neither approach examines how behavioral differentiation across entire populations shapes emergent phenomena like content propagation or network centrality.}
\textit{We address this gap by introducing behavioral traits as an explicit characterization layer that operates alongside identity profiles, enabling heterogeneous, profile-consistent behaviors and linking individual-level dispositions to system-level dynamics.}

\section{Methodology}

\begin{figure*}[ht]
    \centering
    \includegraphics[width=0.9\textwidth]{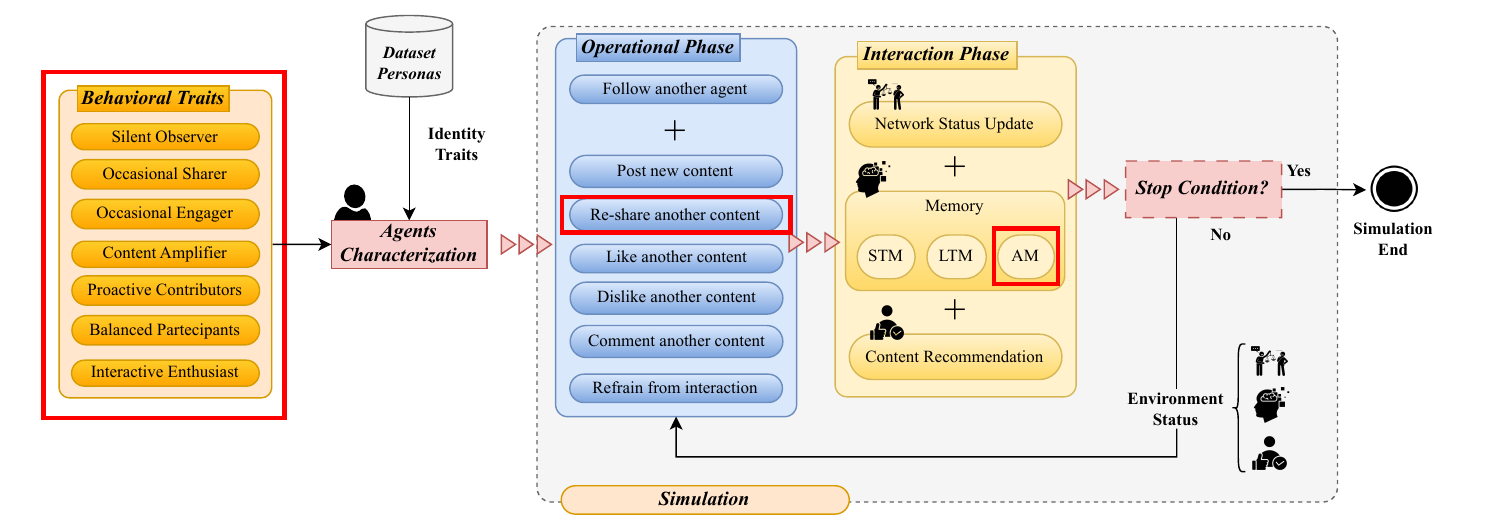}
    \caption{\textbf{Overview of the proposed GABM-based simulation framework.} The framework models large-scale social media dynamics through a population of generative agents, each defined by a two-layer profile of \textit{identity traits} and \textit{behavioral traits}. Agents rely on a three-part \textit{memory unit} (short-term, long-term, and activity memory) to autonomously select and perform platform actions (e.g., posting, following, re-sharing, reacting) until a stop condition is reached. In the figure, the main contributions of this work are highlighted: the introduction of \textit{behavioral traits} and the \textit{activity memory (AM)} component, as well as an extended re-sharing mechanism that allows agents to re-share already re-shared content, enabling content propagation through re-sharing chains.}
    \label{fig:img_framework}
\end{figure*}

This work builds upon an existing GABM framework \cite{orlando2026validating} for social media simulation. The original framework provides the foundational architecture for generative agent-based simulations, including agent memory structures, reasoning modules, and basic interaction mechanisms. Our contribution extends this framework in two critical ways: \textit{(i)} introducing behavioral traits as an explicit layer of agent characterization that regulates participation patterns across the platform's action space, and \textit{(ii)} enabling content propagation chains by allowing re-shared content to be further recommended and amplified. Figure~\ref{fig:img_framework} provides a visual representation of the extended framework.

\subsection{Generative Agents} \label{sec:generative_agents}

\subsubsection{\textbf{Agent Profile.}} \label{sec:agent_profile}

Each agent is characterized by two complementary layers that together define both \textit{who} they are and \textit{how} they act: \textit{(i)} \textbf{identity traits} and \textit{(ii)} \textbf{behavioral traits}. See Appendix~\ref{app:agent_example} for an illustrative example.

\paragraph{Identity Traits.} 
Identity traits provide agents with distinct narrative profiles drawn from the FinePersonas dataset\footnote{\label{fn:finepersonas}\url{https://huggingface.co/datasets/argilla/FinePersonas-v0.1}}. {\color{black}These traits define \textit{who} agents are by encoding background information, interests, and social roles. While identity traits establish an agent's thematic orientation and social identity, they do not specify \textit{how} agents act. This limitation led to homogeneous behaviors in prior work, predominantly content generation \cite{ferraro2024agent, composta2025simulating}.}

\paragraph{Behavioral Traits.} 
Our key contribution is the introduction of behavioral traits as an explicit characterization layer that regulates \textit{how} agents act. These traits instantiate archetypal participation patterns grounded in empirical user typologies. We operationalize seven behavioral traits:

\begin{itemize}[leftmargin=*, itemsep=2pt]

    \item \textbf{Silent Observer (SO):} Fully passive users who do not post or interact except under strong external triggers. This profile aligns with classic "lurker" behavior widely reported in online communities \cite{akar2018user, mut_typology, brandtzaeg2011typology}.

    \item \textbf{Occasional Sharer (OS):} Low-activity users who selectively re-share content but neither create original posts nor engage through comments. This profile reflects sporadic participation patterns observed in social media, where users contribute primarily by amplifying selected content rather than producing or discussing it \cite{lcp_segmentation, mut_typology, temporal_posting_behavior}.

    \item \textbf{Occasional Engager (OE):} Agents with minimal reactive behavior who sporadically like, dislike, or comment on content. This profile corresponds to low-engagement interaction patterns documented in empirical studies of online participation, where users contribute primarily through lightweight reactions \cite{brandtzaeg2011typology, temporal_posting_behavior}. In prior simulation-based models, similar agents exhibit reduced interaction probabilities, providing background engagement that reinforces content visibility \cite{murdock2024agent}.

    \item \textbf{Balanced Participant (BP):} Moderately active users who maintain an equilibrium between original content production and re-sharing. This profile aligns with users who sustain platform activity by contributing across multiple interaction modalities \cite{brandtzaeg2011typology}. Both empirical analyses and agent-based models identify comparable intermediate activity clusters, characterized by balanced probabilities of posting and re-sharing \cite{temporal_posting_behavior, murdock2024agent}.
    
    \item \textbf{Content Amplifier (CA):} Agents oriented toward information propagation through frequent re-sharing and supportive reactions, with limited original content production. This profile reflects amplification-oriented users identified in empirical diffusion studies and agent-based models, including super-spreaders and high-propagation agents \cite{murdock2024agent, temporal_posting_behavior}.

    \item \textbf{Proactive Contributor (PC):} Users focused on generating original content, resembling content creators or opinion leaders consistently observed in online communities \cite{akar2018user, murdock2024agent, temporal_posting_behavior}.
    
    \item \textbf{Interactive Enthusiast (IE):} Highly reactive users who intensively engage with others’ content through likes, dislikes, and comments, while rarely posting or re-sharing. This profile corresponds to \textit{socializers} and highly engaged users described in multiple user typologies \cite{mut_typology, brandtzaeg2011typology, lcp_segmentation}, as well as to high-interaction agents identified in prior simulation studies \cite{murdock2024agent}.
\end{itemize}

{\color{black}These behavioral traits are not mutually exclusive categories with rigid boundaries. Crucially, behavioral prompts are not encoded as deterministic constraints on the agent's action space but rather as soft dispositional priors that the LLM interprets against the agent's identity, recent activity, and the current simulation context. As a result, while each trait exhibits a predominant behavioral orientation, agents may occasionally perform actions outside their primary tendency, or even actions explicitly discouraged by their assigned trait, when the simulation context warrants it. This soft-rule interpretation preserves the capacity for context-sensitive deviation that distinguishes generative agents from rule-based ones, and it allows the framework to reproduce the behavioral variability observed in real users, who exhibit predominant participation styles while occasionally departing from them. We validate this property empirically in Section~\ref{sec:RQ1}.}

Both identity and behavioral traits are embedded directly into each agent's system prompt, ensuring consistent guidance throughout decision-making. This dual-layer architecture is critical: identity traits maintain narrative coherence while behavioral traits prevent the collapse into homogeneous participation patterns observed in prior work. {\color{black} We assume both traits remain fixed throughout the simulation\footnote{{\color{black}While real users' participation styles and interests may evolve over time, modeling such adaptive dynamics falls outside the scope of this study. Instead, we treat identity and behavioral traits as stable characterization dimensions and assess the extent to which agents remain faithful to their assigned profiles.}}.}

See Appendix~\ref{app:behav_traits_prompt} for the full set of behavioral trait prompts.

Beyond their characterization layers, agents maintain dynamic social connections. While identity and behavioral traits remain fixed throughout the simulation, the network of following relationships evolves as agents autonomously decide whom to follow based on their ongoing experiences and interactions.

\subsubsection{\textbf{Memory Unit.}}

The original framework \cite{orlando2026validating} equips each agent with a two-layer memory architecture designed to balance recent and significant interactions:

\paragraph{Short-Term Memory (STM)} stores recently encountered content along with engagement metadata (number of received re-shares, likes, dislikes, and comments). A decay mechanism removes less relevant items over time to prevent information overload.

\paragraph{Long-Term Memory (LTM)} retains high-impact content identified through periodic evaluations of the STM. Content is transferred based on engagement scores derived from re-shares, likes, dislikes, and sentiment analysis of associated comments, ensuring that only the most influential items persist.

{\color{black}\paragraph{Activity Memory (AM)} Our extension introduces a third memory component which maintains a record of the agent's recent actions within the platform's action space. For each agent we maintain a fixed-size vector 
\[
\mathbf{a} = (a_{\text{post}}, a_{\text{reshare}}, a_{\text{interact}}, a_{\text{inactive}}) \in \mathbb{N}^4,
\]
where each component $a_k$ represents the number of iterations since the agent last performed action type $k$ (posting original content, re-sharing, interacting via likes/dislikes/comments, or remaining inactive). At each iteration $t$, when the agent selects action $k^\ast$, the corresponding counter is reset to zero ($a_{k^\ast} \leftarrow 0$), while all other counters are incremented by one ($a_k \leftarrow a_k + 1$ for $k \neq k^\ast$). The values of \textbf{a} are then serialized into a structured textual report describing \textit{(i)} the most recent action, \textit{(ii)} the number of iterations elapsed since each action type was last executed, and \textit{(iii)} whether any actions have never been executed. This mechanism is critical for enabling behavioral trait consistency: without awareness of their own action history, agents cannot align their decisions with their assigned behavioral profiles. For instance, a Balanced Participant that has posted several times in a row can shift toward re-sharing to rebalance its activity, while an Occasional Engager that has been inactive for several iterations can identify the gap and resume sporadic interactions.}

All three memory components—STM, LTM, and AM—are all initialized as empty, enabling agents’ memories to evolve exclusively through activities during the simulation and thus providing an unbiased basis for assessing the effects of behavioral traits.

\subsubsection{\textbf{Reasoning Module.}}

Agent decision-making is implemented through a prompt-based interaction with an LLM, which emulates user behavior at each simulation step. We followed the original framework \cite{orlando2026validating} without modifications: at each iteration, the module produces a \emph{Choice–Reason–Content} triplet. The \emph{Choice} component specifies the action selected from the platform's action space (e.g., posting original content, re-sharing another agent's content, following another agent, or remaining inactive). The \emph{Reason} component provides the rationale motivating the action selection. The \emph{Content} component contains the generated material associated with the action, such as the text of an original post, a comment on another agent's content, or the identifier of content to re-share or interact with. When the agent chooses inactivity, this field remains empty.
To mitigate potential hallucinations, agents are automatically re-prompted whenever the generated response does not comply with the expected format or refers to non-existent content. In particular, agents are required to select one item from the set of recommended content, and if the selection is invalid or does not correspond to any available item, the prompt is reissued until a valid choice is made.

\subsection{Simulation Workflow} \label{subsec:sim_workflow}

The simulation follows an iterative workflow articulated into three phases. During the \textbf{Initialization Phase}, agents are configured with distinct personalities and behavioral traits, and their memory components (STM, LTM, and AM) are initialized as empty. The \textbf{Operational Phase} immerses agents in an environment resembling a social media platform, where they autonomously make decisions at each iteration. The \textbf{Interaction Phase} updates the environment state based on agents' actions, processes engagement metrics, and prepares the context for the next iteration.

\paragraph{\textbf{Agent Decision-Making.}}

We followed the workflow proposed in the original framework \cite{orlando2026validating}: at each iteration of the Operational Phase, the input prompt to each agent's Reasoning Module is constructed by incorporating information a real user would likely consider before acting: \textit{(i)} feedback on previously published content, retrieved from the agent's memory unit (STM and LTM) in terms of re-shares, likes, dislikes, and comments; \textit{(ii)} a summary of recent activity patterns from the Activity Memory, describing what actions the agent has recently performed and when each action type was last executed; \textit{(iii)} a personalized feed of recommended posts from other agents available for potential interaction; and \textit{(iv)} the list of available actions the agent can perform—posting new content, re-sharing, liking, disliking, commenting, following another agent, or remaining inactive. {\color{black} Following the original framework \cite{orlando2026validating}, we consider two recommendation strategies: a \emph{preference-based} strategy, where candidate posts and the agent’s content are mapped into a shared embedding space and the top-$k$ posts by cosine similarity are returned, so that the recommended content aligns with the agent's interests; and a \emph{random} strategy, where $k$ posts are sampled uniformly from the available pool.} The Operational Phase concludes once each agent’s Reasoning Module produces the \emph{Choice–Reason–Content} triplet. This workflow continues iteratively until a predefined stop condition (e.g., based on number of iterations) is reached.

\paragraph{\textbf{Content Propagation Mechanism.}}

A critical extension we introduce is the modification of the recommendation system to enable propagation chains. Unlike the original framework \cite{orlando2026validating}, where re-shared content was not propagated to the feeds of followers of the re-sharing agent, our design addresses this limitation by extending the recommendation mechanism to include re-shared content. As a result, when an agent re-shares content, that re-shared item becomes visible to the re-sharing agent's followers and can itself be recommended in their personalized feeds. This allows re-shared items to be reacted to and re-shared further, enabling the formation of propagation chains \cite{vosoughi2018spread,goel2016structural,migliorini2023tracking}. This modification enables the study of amplification dynamics and supports the evaluation of behavioral traits specifically designed for diffusion (e.g., Content Amplifiers).

\section{Experiments}\label{sec:experiments}

\subsection{Experimental Setup}

To systematically assess the impact of behavioral traits, personality representations, and content exposure mechanisms, we conducted simulations under four complementary configurations:

\begin{itemize}
    \item \texttt{FullModel} (Behavioral Traits + Preference-based Recommendation): The complete framework combining FinePersonas-based identity profiles, the proposed behavioral traits, Activity Memory, and a preference-based recommender aligned with agents' topical interests. This represents our proposed approach.

    \item \texttt{IdentityOnly}: Agents characterized solely by FinePersonas-based identity profiles without the proposed behavioral traits, exposed to the same preference-based recommender. This ablation refers to the original implementation of the GABM framework \cite{ferraro2024agent}.

    \item \texttt{RandomRecommendation}: The \texttt{FullModel} but with content exposure governed by a random recommender rather than preference-based recommender. This configuration isolates the effect of recommendation strategy on propagation dynamics while maintaining behavioral trait regulation.

    %\item \texttt{PsychometricTraits}: Agents retain FinePersonas-based identity profiles and preference-based recommendation, but explicit behavioral traits are replaced by psychometric descriptors from the OCEAN personality model \cite{mccrae1992introduction}. Following \cite{huang2026designing}, each factor is instantiated in high and low variants via the agent's system prompt. This configuration tests whether psychometric personality traits alone can induce behavioral differentiation comparable to explicit behavioral traits.

    \item {\color{black}\texttt{OCEANTraits}: Since prior research has reported associations between OCEAN traits and social media engagement behaviors such as posting frequency, content sharing, and interaction patterns \cite{correa2010interacts, kosinski2014manifestations}, agents in this configuration retain FinePersonas-based identity profiles and preference-based recommendation, paired with personality descriptors from the OCEAN model \cite{mccrae1992introduction}. Following \cite{huang2026designing}, each factor is instantiated in high and low variants. This configuration examines whether OCEAN descriptors are sufficient to elicit heterogeneous participation patterns in generative agents.}
\end{itemize}

For the three configurations using identity traits, we instantiate agent populations by sampling identity profiles from FinePersonas\footref{fn:finepersonas} across four thematic domains: Healthcare, Technology, Religion, and Music. Each domain contains 35 distinct personas, for a total of 140 distinct identity profiles. In configurations with explicit behavioral traits, each identity profile is paired with one of the seven behavioral profiles defined in Section~\ref{sec:agent_profile}, yielding 245 agents per domain and 980 agents total. This balanced design ensures that observed differences are attributable to behavioral trait assignment rather than demographic composition or topical distribution. For the \texttt{IdentityOnly} configuration, we instantiate the same 980 FinePersonas profiles without behavioral trait assignment.

All simulations ran for 25 iterations, {\color{black}following the stop condition adopted in the original framework \cite{orlando2026validating}. This duration is kept short because extending the simulation would require accounting for the possibility that agents' traits evolve dynamically, which falls outside the scope of this study.} We repeated simulations using two LLMs: Llama 3 70B and Gemma 3 27B. Together, these simulations allow us to disentangle the effects of behavioral traits, personality representations, and content exposure mechanisms on emergent behavioral patterns and network-level outcomes. Implementation details are provided in Appendix \ref{app:implementation}.

\begin{figure*}[t]
     \centering
     \subfloat[][]{\includegraphics[width=.27\linewidth]{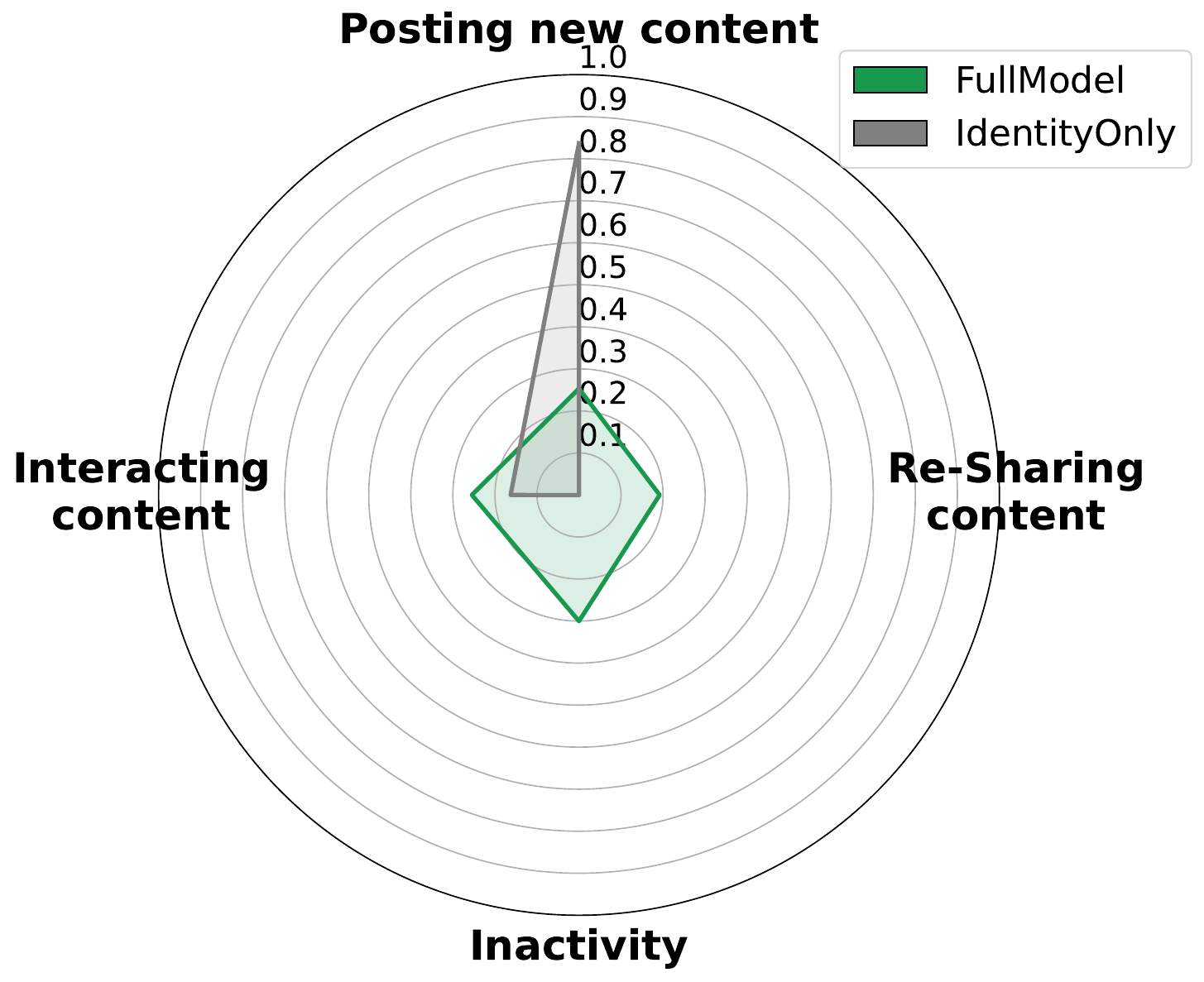}\label{fig:radar_all}}
     \subfloat[][]{\includegraphics[width=.255\linewidth]{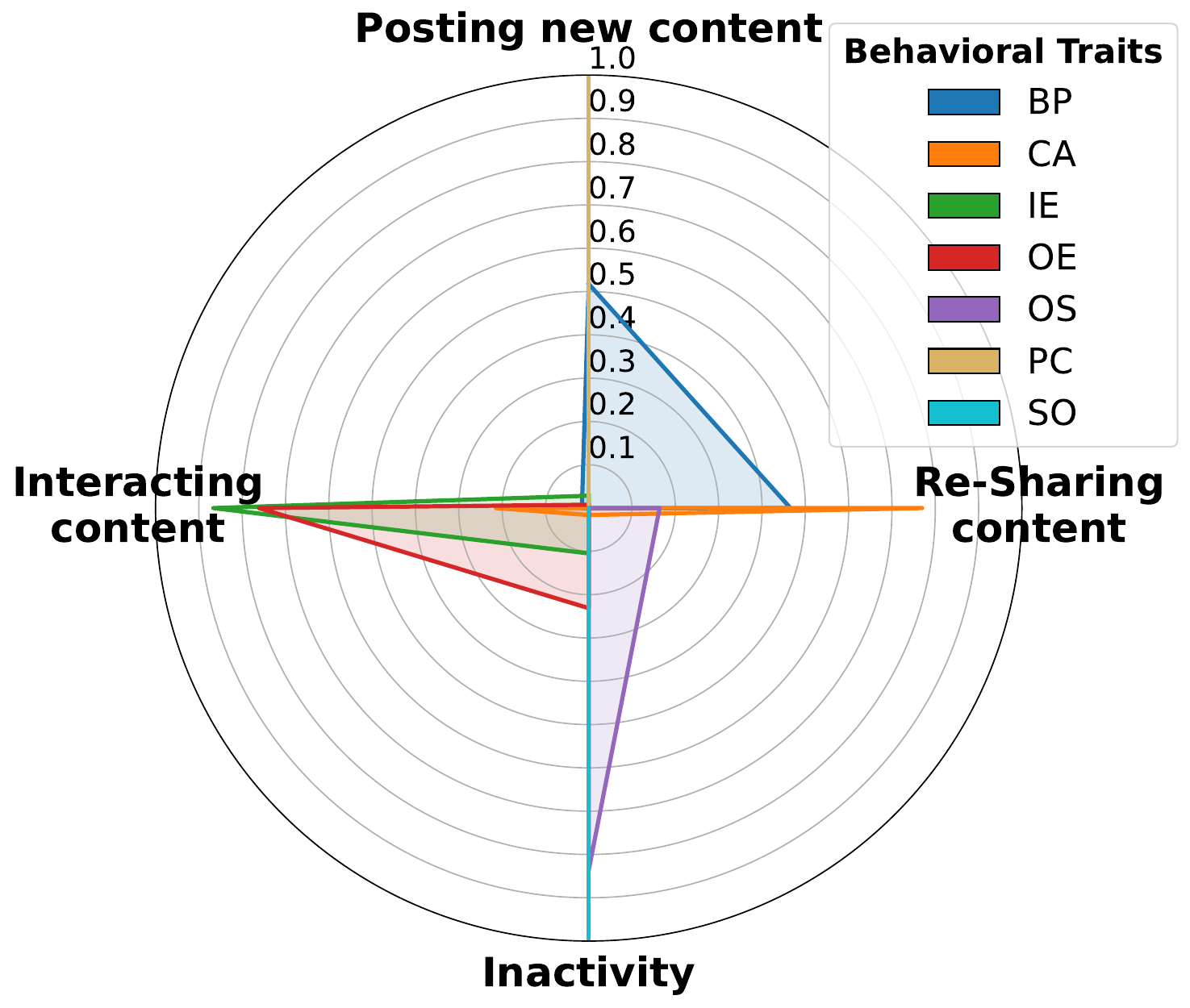}\label{fig:radar_sovrapposti}}
     \subfloat[][]{\includegraphics[width=.255\linewidth]{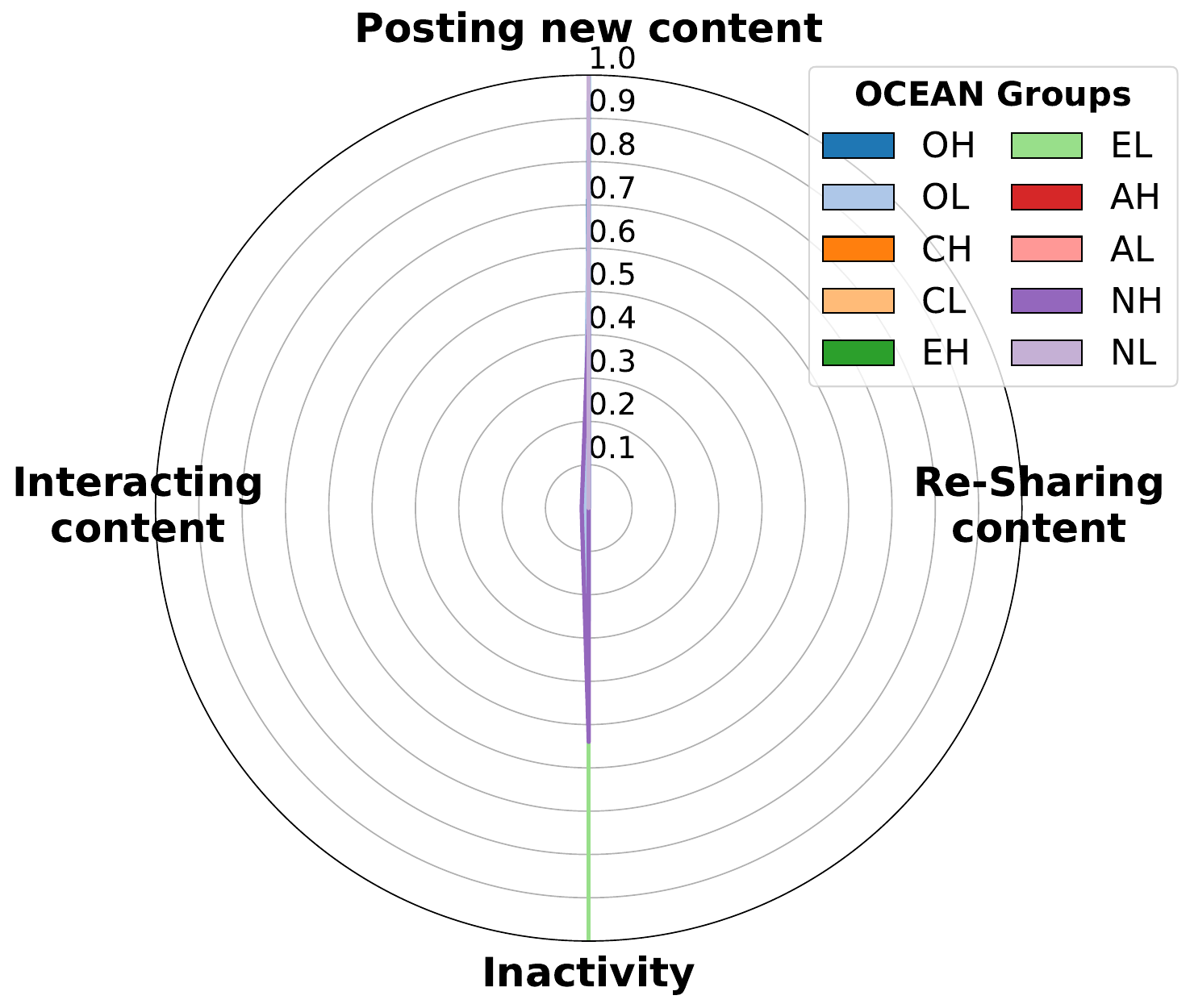}\label{fig:ocean_radar}}
     \subfloat[][]{\includegraphics[width=.23\linewidth]{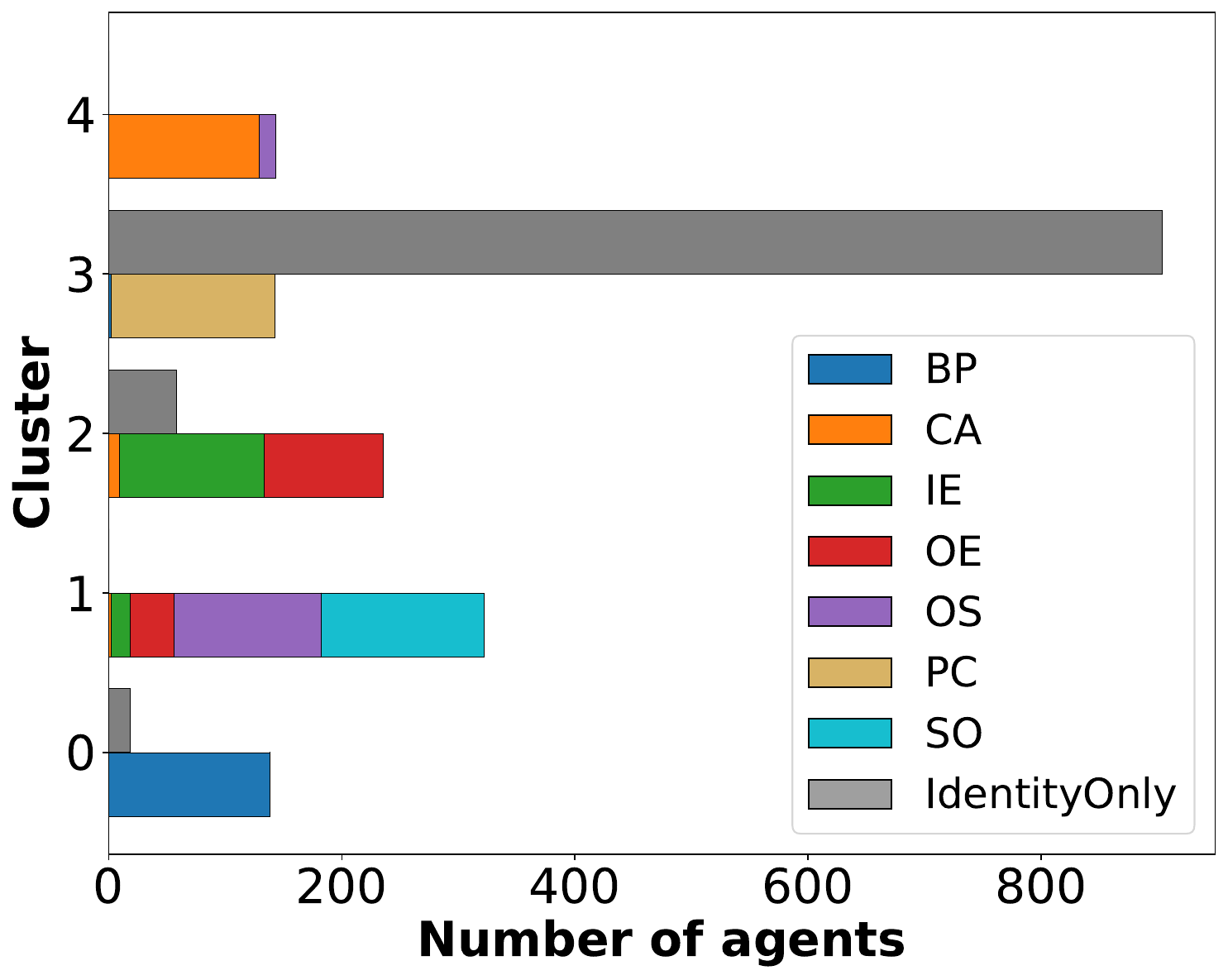}\label{fig:cluster_distributions}}
      \caption{(a) Average action probabilities in the \texttt{FullModel} and \texttt{IdentityOnly} (behavioral traits disabled) configurations. (b) Average action probabilities across the seven behavioral traits. (c) Average action probabilities across the ten OCEAN traits. (d) Comparison of cluster distributions in the \texttt{FullModel} and \texttt{IdentityOnly} configurations. With traits enabled, agents form diverse, profile-consistent clusters, while without traits, most collapse into a single cluster.}
     \label{fig:rq1_fig}
\end{figure*}

\subsection{Behavioral Traits Alignment (RQ1)}  \label{sec:RQ1}

Here, we investigate whether generative agents exhibit action patterns that are consistent with their assigned behavioral traits. Figure~\ref{fig:radar_sovrapposti} illustrates the average probability of performing each action type—$P_{\text{post}}$, $P_{\text{re-share}}$, $P_{\text{interact}}$, $P_{\text{inactive}}$—across behavioral profiles (detailed probability distributions are provided in Appendix~\ref{app:action_prob_distr}) for simulations with Llama 3 agents. The results demonstrate strong profile consistency: Proactive Contributors dedicate nearly all activity to posting original content ($P_{\text{post}} \approx 1$), while Silent Observers remain systematically inactive ($P_{\text{inactive}} \approx 1$). Balanced Participants maintain their intended equilibrium, dividing effort between posting (51.76\%) and re-sharing (46.58\%). Content Amplifiers enact their propagation-oriented role primarily through re-sharing (76.96\%), supplemented by occasional interactions (21.43\%) that complement their amplification function. Engagement-focused profiles exhibit distinct patterns: Interactive Enthusiasts concentrate overwhelmingly on interactions—likes, dislikes, and comments (86.67\%)—representing the highest engagement rate across all profiles. Occasional Engagers similarly prioritize interactions (76.10\%) but at substantially lower frequencies, consistent with their sporadic design. Both OE and Occasional Sharers display elevated inactivity levels, with OS selectively re-sharing during active periods while remaining passive otherwise. Similar patterns emerge for simulations with Gemma 3 agents; see Appendix~\ref{app:gemma_behav} for further details.

To further quantify this heterogeneity, we compare the \texttt{FullModel} configuration against both the \texttt{IdentityOnly} and the {\color{black}\texttt{OCEANTraits}} configurations.

\paragraph{\texttt{FullModel} vs \texttt{IdentityOnly}.}

%Figure \ref{fig:radar_all} reports the average probabilities of taking each action across both configurations for simulations with Llama 3 agents. In the \texttt{IdentityOnly} setting (grey), agents exhibit a pronounced bias toward generating new content (83.665\%), while largely disregarding re-sharing (0.004\%), interactions (16.269\%), and inactivity (0.006\%). This homogeneity, previously observed in prior GABM research \cite{ferraro2024agent}, indicates that personality profiles alone fail to diversify participation patterns. In contrast, the \texttt{FullModel} configuration (green) produces a balanced representation across the full action space, with posting (25.310\%), re-sharing (19.188\%), interactions (25.478\%), and inactivity (30.024\%) all contributing to heterogeneous behavioral patterns. {\color{black}To assess the statistical significance of these differences, we compared the per-agent action probabilities between the two configurations using a paired t-test and a Wilcoxon signed-rank test. The differences are statistically significant under both tests for posting, re-sharing, and inactivity ($p<0.05$). The only exception is the interaction probability, where the t-test detects a significant increase in the \texttt{FullModel} configuration ($p<0.05$) but the Wilcoxon test does not ($p>0.05$).}

{\color{black}We compare per-agent action probabilities between the two configurations using a Wilcoxon signed-rank test. Figure~\ref{fig:radar_all} reports the results for simulations with Llama 3 agents. Posting shifts from $83.665\%$ in \texttt{IdentityOnly} to $25.310\%$ in \texttt{FullModel} ($p{=}1.09{\times}10^{-140}$), re-sharing from $0.004\%$ to $19.188\%$ ($p{=}2.98{\times}10^{-65}$), and inactivity from $0.006\%$ to $30.024\%$ ($p{=}2.42{\times}10^{-64}$). The only exception is the interaction probability, which increases from $16.269\%$ to $25.478\%$: the Wilcoxon test does not detect a significant difference ($p{=}0.271$). The homogeneity of the \texttt{IdentityOnly} configuration, previously observed in GABM research \cite{ferraro2024agent}, indicates that identity traits alone fail to diversify participation patterns. In contrast, the \texttt{FullModel} configuration (green) produces a balanced representation across the full action space.}

To quantify the heterogeneity of the \texttt{FullModel} configuration, we clustered agents based on their average action probabilities. Clustering validity analysis using the Elbow Method and Silhouette Score indicated $k=5$ as optimal. Figure~\ref{fig:cluster_distributions} shows that resulting clusters align closely with intended behavioral traits. Cluster 3 comprises exclusively Proactive Contributors (PC), while Cluster 0 contains only Balanced Participants (BP). Cluster 1 aggregates passive agents: all Silent Observers (SO) reside here (43.5\% of cluster membership), alongside Occasional Sharers (OS) and Occasional Engagers (OE) exhibiting sporadic inactivity. Cluster 2 groups interaction-oriented agents—Interactive Enthusiasts (IE), Content Amplifiers (CA), and Occasional Engagers (OE)—while Cluster 4 merges Content Amplifiers and Occasional Sharers, both characterized by re-sharing behaviors.

We then projected agents from the \texttt{IdentityOnly} configuration onto these fixed centroids to test whether different behavioral profiles emerged without explicit trait assignment. As Figure~\ref{fig:cluster_distributions} demonstrates, the distribution collapsed almost entirely into Cluster~3 (the Proactive Contributor cluster), which comprises 904 users, leaving Clusters~0 and~2 sparsely populated, comprising 18 and 58 users, respectively, and Clusters~1 and~4 remain empty. This confirms the strong content-generation bias of agents initialized without behavioral profiles.

These results demonstrate that behavioral traits successfully constrain agent decision-making, yielding heterogeneous yet profile-consistent behaviors, whereas their absence leads to homogeneous content-generation patterns.

\paragraph{\texttt{FullModel} vs {\color{black}\texttt{OCEANTraits}}.}

%We compare the \texttt{FullModel} configuration with the {\color{black}\texttt{OCEANTraits}} configuration to assess whether psychometric personality traits alone are sufficient to induce heterogeneous action patterns in generative agents.

Figure \ref{fig:ocean_radar} reports the average probabilities of making each action (post, re-share, interact, and inactive) across the ten OCEAN profiles (five traits, each with high and low variants \cite{huang2026designing}). The radar plot reveals strong homogeneity: regardless of the specific OCEAN trait, agents predominantly concentrate their behavior on content generation and inactivity, while re-sharing and interaction probabilities remain close to zero.

These results indicate that the only difference between the \texttt{IdentityOnly} and the {\color{black}\texttt{OCEANTraits}} configurations (see Figures \ref{fig:radar_all} and \ref{fig:ocean_radar}) is that the latter yields to higher inactivity for some profiles such as Low Extraversion (EL) or High Neuroticism (NH). This suggests that instantiating agent behavior through OCEAN traits primarily modulates \emph{how much} agents act, rather than \emph{how} they act. In contrast, the \texttt{FullModel} configuration yields diversified and role-consistent engagement patterns across posting, re-sharing, interaction, and inactivity. {\color{black}These results demonstrate that OCEAN traits alone are insufficient to produce differentiated participation roles in social media simulations. Mann-Whitney tests confirm that the differences in action probabilities between the \texttt{OCEANTraits} and the \texttt{FullModel} configurations are statistically significant for all four action types (posting: $p=6.20\times10^{-110}$; re-sharing: $p=2.06\times10^{-106}$; inactivity: $p=5.28\times10^{-8}$; interaction: $p=5.14\times10^{-116}$).} Taken together, these findings indicate that explicit behavioral traits are required to regulate agent decision-making across the action space and to prevent the collapse of agent behavior into uniform content production.

\subsection{Behavioral traits \& Content Propagation (RQ2)}\label{sec:RQ2}

Here, we examine the impact of behavioral traits on content propagation dynamics. As demonstrated in Section~\ref{sec:RQ1}, the \texttt{IdentityOnly} and {\color{black}\texttt{OCEANTraits}} configurations produce virtually no re-sharing activity, preventing the formation of propagation chains entirely. In contrast, explicit behavioral profiling enables agents to engage in re-sharing and interaction actions, thus promoting the potential formation of propagation chains. For this reason, in the following sub-sections, we analyze how content propagates through the simulation for the \texttt{FullModel} configuration and \texttt{RandomRecommendation} configuration.

\subsubsection{\textbf{Content Propagation}}

To analyze how content propagates within the simulation, we introduce a distinction between \textit{first-order actions} and \textit{second-order actions}. First-order actions are engagements directly with original content—re-sharing a post created by its author, or liking, disliking, and commenting on original posts. Conversely, second-order actions are engagements with previously re-shared content—further re-sharing an already re-shared post, or interacting with re-shared items. For example, if Agent A posts original content, Agent B re-shares it, and Agent C subsequently re-shares Agent B's re-share, then Agent B performs a first-order action while Agent C performs a second-order action. This distinction is critical for understanding amplification cascades: first-order actions represent direct engagement with content creators, while second-order actions capture how content spreads beyond its original audience through successive propagation layers. Notably, this analytical framework is enabled by our methodological contribution described in Section~\ref{subsec:sim_workflow}: unlike the original framework \cite{orlando2026validating}, where re-shared content was not propagated to followers' feeds, our design allows re-shared items to appear in personalized recommendations and be further amplified, enabling the formation of propagation chains \cite{vosoughi2018spread, goel2016structural}.

In the \texttt{FullModel} configuration, results show a balanced distribution between the two categories: 59.17\% of actions are first-order, while 40.83\% are second-order. In the case of the \texttt{RandomRecommendation} configuration, results show a slightly more balanced distribution: 51.92\% of actions are first-order, while 48.08\% are second-order.

\begin{figure}[t]
    \centering
    \begin{subfigure}[t]{0.495\columnwidth}
        \centering
        \includegraphics[width=\linewidth]{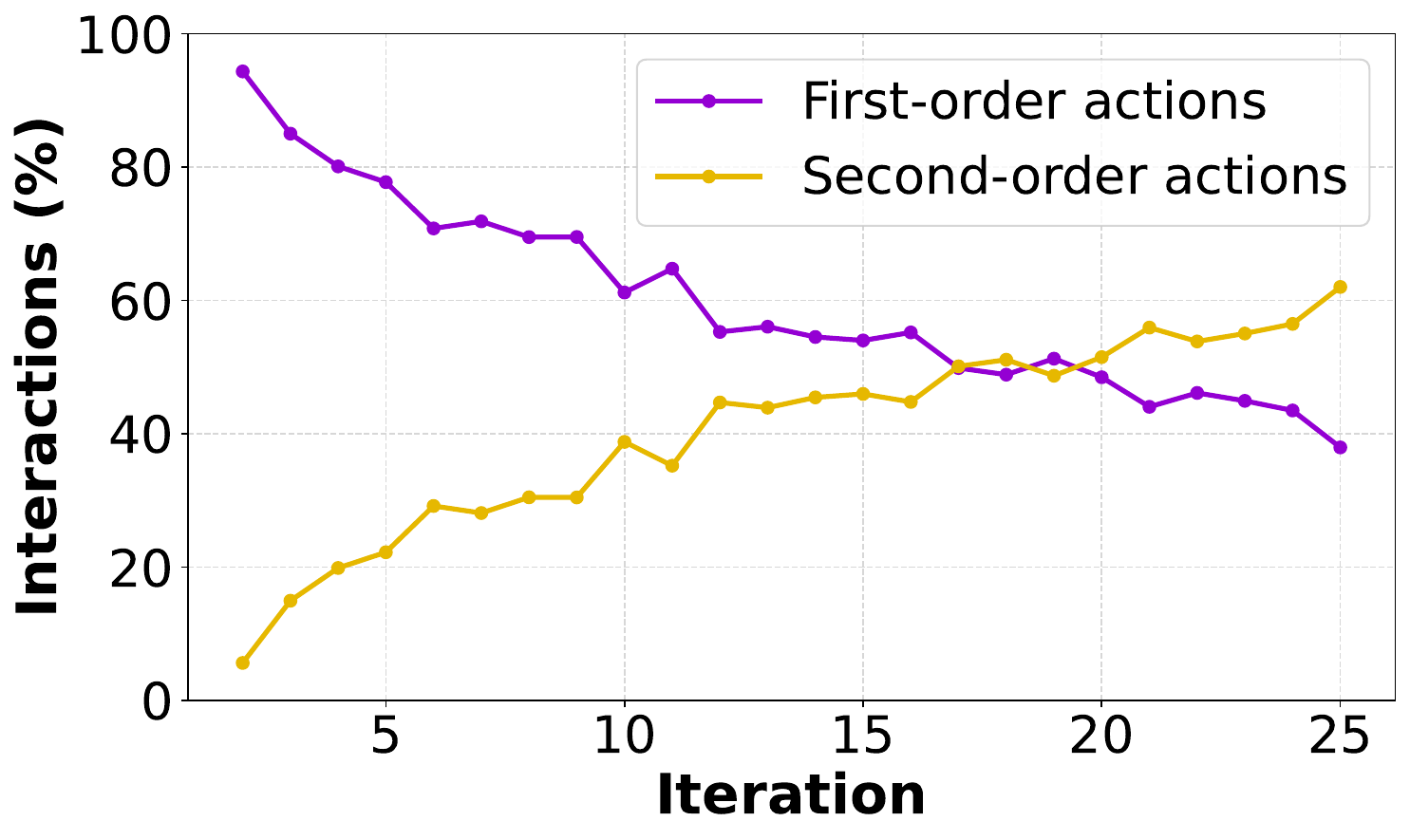}
        \caption{} %First- vs. Second-Order Actions}
        \label{fig:percent_dirette_indirette_tempo}
    \end{subfigure}
    \hfill
    \begin{subfigure}[t]{0.495\columnwidth}
        \centering
        \includegraphics[width=\linewidth]{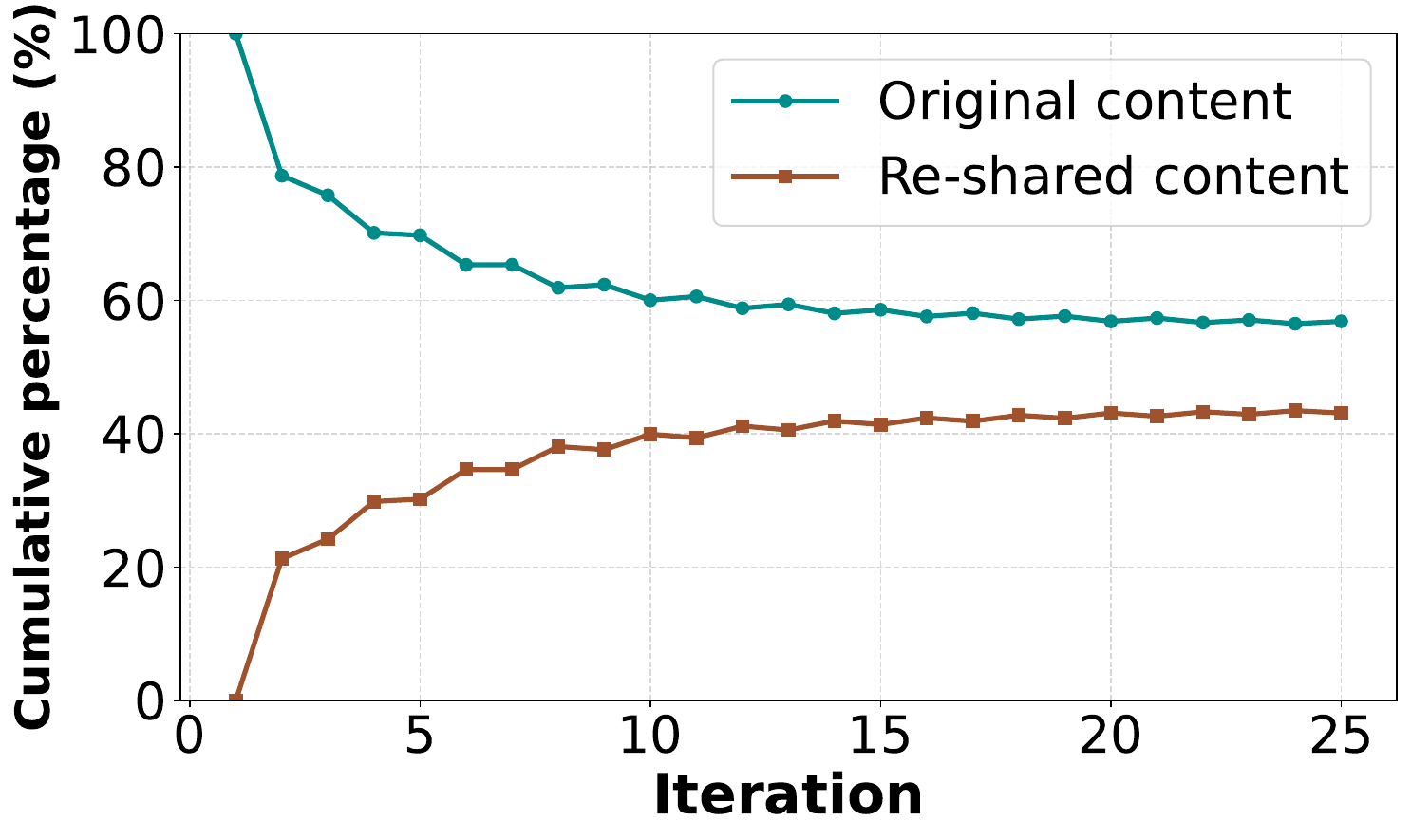}
        \caption{} %Original vs. Re-shared Content}
        \label{fig:percent_originali_condivisi}
    \end{subfigure}
    \caption{Comparison of content production and amplification dynamics for the \texttt{FullModel} configuration, illustrating \textit{(a)} the temporal evolution of first- and second-order actions and \textit{(b)} the cumulative percentage of original and re-shared content.}
    \label{fig:contenuti_interazioni}
\end{figure}

To analyse temporal dynamics of first- and second-order actions, we measure their prevalence for each iteration. Figure~\ref{fig:percent_dirette_indirette_tempo} shows the percentage of first-order (purple) and second-order (yellow) actions across iterations for the \texttt{FullModel} configuration. We report the same results for the \texttt{RandomRecommendation} configuration in Appendix \ref{app:Behavioral_Random_dynamics} since they convey similar findings. Initially, only first-order actions are observed, as agents engage directly with the limited pool of original content available in the early simulation stages. As the simulation progresses, the proportion of first-order actions steadily decreases while second-order actions increase correspondingly. By iteration 19, second-order actions surpass first-order actions, indicating that agents are increasingly engaging with already re-shared content rather than exclusively interacting with original posts. 

This trend suggests the formation of content propagation chains but also raises a potential concern: if agents primarily engage with re-shared content, the simulation risks saturation—a scenario where original content production declines or ceases entirely, leaving only recursive amplification of existing material. Figure~\ref{fig:percent_originali_condivisi} addresses this concern by tracking the temporal evolution of original versus re-shared content production. As expected, there are 100\% original content at the beginning of the simulation, since no prior posts are available for re-sharing. Over subsequent iterations, the proportion of re-shared content increases as agents encounter and amplify existing posts. However, this growth does not lead to the collapse of original content production. Instead, the distribution stabilizes at a plateau where both original and re-shared content coexist in equilibrium.

\subsubsection{\textbf{Propagation Chain Analysis}}

We define a propagation chain as a sequence of successive re-shares originating from a single original post. The chain length represents the number of re-sharing events, with length 2 indicating original content re-shared once, length 3 indicating that re-shared content was further propagated, and so forth. We measure chain length by tracing each re-share back to its original source, capturing the depth of content diffusion through the network.

Table~\ref{tab:chain_lengths} reports the distribution of chain lengths across both recommendation configurations. In the \texttt{FullModel} configuration, the average chain length is 2.53, with the majority of chains terminating at length 2 (2,346 cases) or length 3 (868 cases). Nonetheless, substantially longer chains emerge, reaching a maximum depth of 10, indicating that certain content achieves deep diffusion through multiple propagation layers. The \texttt{RandomRecommendation} configuration exhibits a similar distribution, but with a significantly higher average chain length of 2.79 (Mann-Whitney test, {\color{black}$p=2.71\times10^{-36}$}), though the maximum depth decreases marginally to 9.

This difference reflects distinct diffusion mechanisms induced by recommendation strategies. Under preference-based recommendation, the distribution is highly skewed toward short chains (predominantly length 2), with 64.82\% of chains terminating after a single re-share. This reflects dynamics where agents repeatedly engage with content closely aligned to their topical interests, producing numerous isolated re-shares but limiting overall diffusion depth. The concentration of engagement within topical niches creates high-volume but shallow cascades, as content rarely crosses thematic boundaries to reach broader audiences. Conversely, random content exposure increases the likelihood of agents encountering previously re-shared posts from diverse topics, occasionally triggering additional re-shares that extend propagation depth. This produces a flatter distribution with more mid-length chains (lengths 3–5: 45.68\% under random vs. 33.84\% under preference-based) but fewer extremely short chains (51.99\% at length 2 under random vs. 64.82\% under preference-based). The diversification of content encounters enables deeper propagation, as re-shared content reaches agents outside the original topical community, increasing opportunities for successive amplification layers.

We further examine whether propagation dynamics vary across thematic domains. Each agent's personality is associated with one of four topics (Healthcare, Technology, Religion, Music), and a chain's topic is determined by its original post. The average chain length across topics shows minimal variation: healthcare (2.47), music (2.54), religion (2.66), and technology (2.45). The distribution of chains is similarly balanced, with healthcare accounting for 27.5\%, music 23.1\%, religion 28.8\%, and technology 25.6\% of total chains. These findings indicate that amplification dynamics are largely independent of topical content, a result consistent with the balanced initialization of agents across domains (245 agents per topic) and reinforcing that behavioral traits—rather than content themes—drive propagation patterns.

\begin{table}[t]
    \centering
    \caption{Frequencies and relative percentages of propagation chain lengths across the \texttt{FullModel} and \texttt{RandomRecommendation} configurations.}
    \label{tab:chain_lengths}
    \resizebox{0.9\columnwidth}{!}{%
    \begin{tabular}{c c c}
        \hline
        \textbf{Chain length} & \texttt{FullModel} & \texttt{RandomRecommendation} \\
        \hline
        2  & 2346 (64.82\%) & 2172 (51.99\%) \\
        3  & 868 (23.98\%)  & 1158 (27.72\%) \\
        4  & 257 (7.1\%)  & 528 (12.64\%)  \\
        5  & 100 (2.76\%)  & 210 (5.03\%)  \\
        6  & 29 (0.8\%)   & 83 (1.99\%)   \\
        7  & 11 (0.3\%)   & 24 (0.57\%)   \\
        8  & 5 (0.14\%)   & 2 (0.05\%)    \\
        9  & 2 (0.06\%)   & 1 (0.02\%)    \\
        10 & 1 (0.03\%)    & 0 (0.0\%)    \\
        \hline
        \textbf{Total} & 3619 & 4178\\
        \hline
    \end{tabular}%
    }
\end{table}

\subsubsection{Propagation Chain \& Behavioral Profiles}

 To examine which behavioral profiles drive content diffusion at different stages, we analyze the composition of propagation chains by position. Each position represents a layer in the cascade: position 0 corresponds to the original content creator, position 1 to the first re-share, position 2 to subsequent re-shares, and so forth. We measure the percentage contribution of each behavioral profile at every position across all observed chains.

\begin{figure}[t]
    \centering
    \includegraphics[width=\columnwidth]{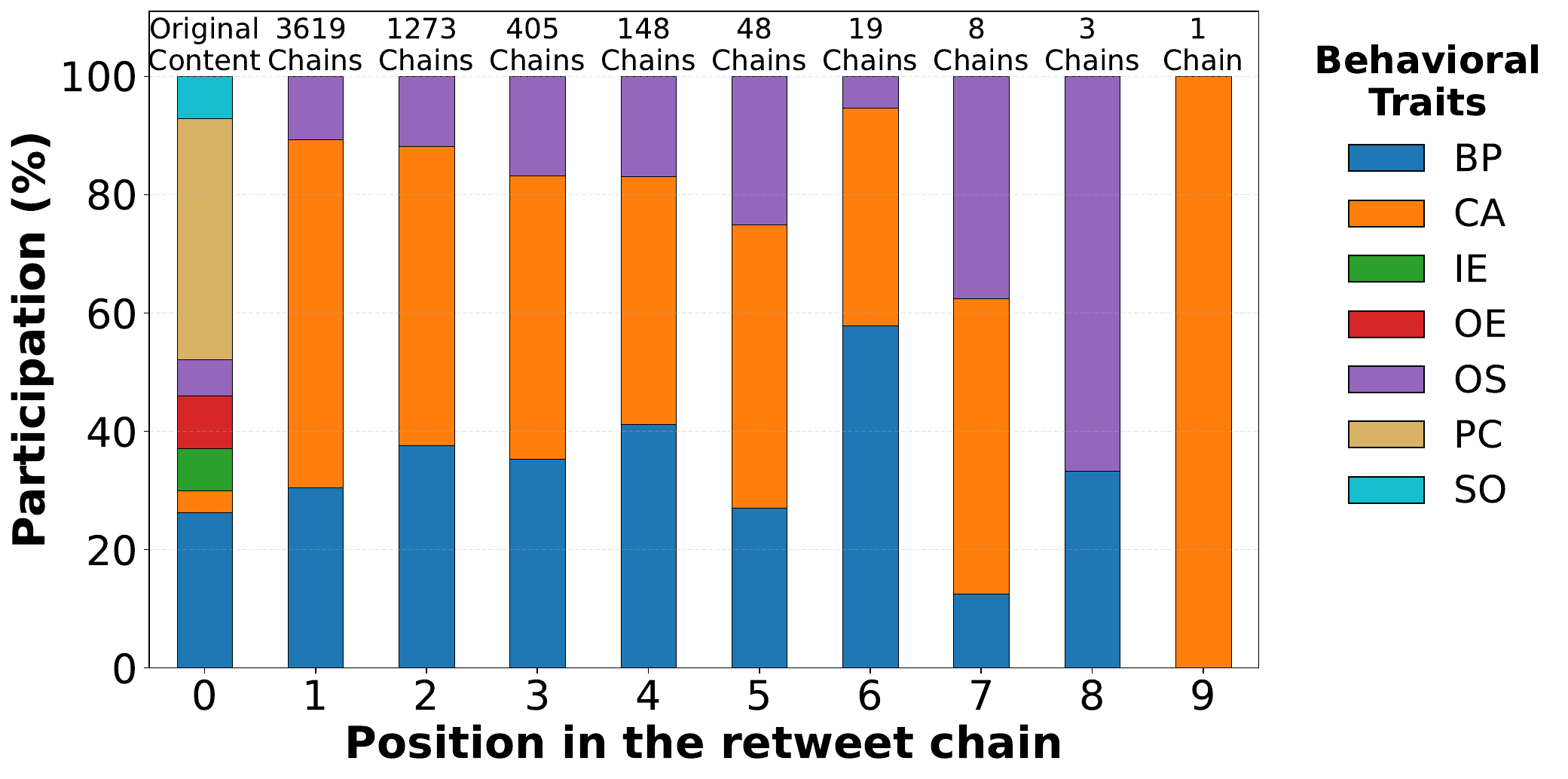}
    \caption{Distribution of behavioral traits across positions in propagation chains. Position 0 corresponds to the creation of original content, while subsequent positions represent successive re-shares. Bars indicate the percentage of agents from each behavioral trait contributing at each stage.}
    \label{fig:perc_allPos}
\end{figure}

Figure~\ref{fig:perc_allPos} illustrates how group composition evolves across chain positions for the \texttt{FullModel} configuration. At position 0—the creation of original content—nearly all profile types appear, since agents in the first simulation iteration can either post or remain inactive, while other actions (re-sharing, commenting, liking) depend on pre-existing content. Nonetheless, Proactive Contributors (PC) and Balanced Participants (BP) dominate this initial stage accounting for approximately 40\% and 25\%, respectively, of the  agents in this position. 

As chains extend beyond the original post, a stark pattern emerges: subsequent positions are almost exclusively occupied by resharing-oriented profiles. Content Amplifiers (CA), Balanced Participants (BP), and Occasional Sharers (OS) collectively account for over 90\% of all re-sharing activity at positions 1 and beyond, while Proactive Contributors, Interactive Enthusiasts, and Occasional Engagers contribute minimally to further diffusion. % This concentration intensifies at deeper chain positions, where Content Amplifiers alone represent the majority of re-sharers.

Taken together, these results show that behavioral traits are necessary to enable content propagation: without them, no diffusion chains emerge. When enabled, propagation dynamics arise from the interplay between amplification-oriented profiles (Content Amplifiers, Balanced Participants, Occasional Sharers) that extend content reach.

\begin{figure}[t]
     \centering
     \subfloat[][]
     {\includegraphics[width=.51\columnwidth]{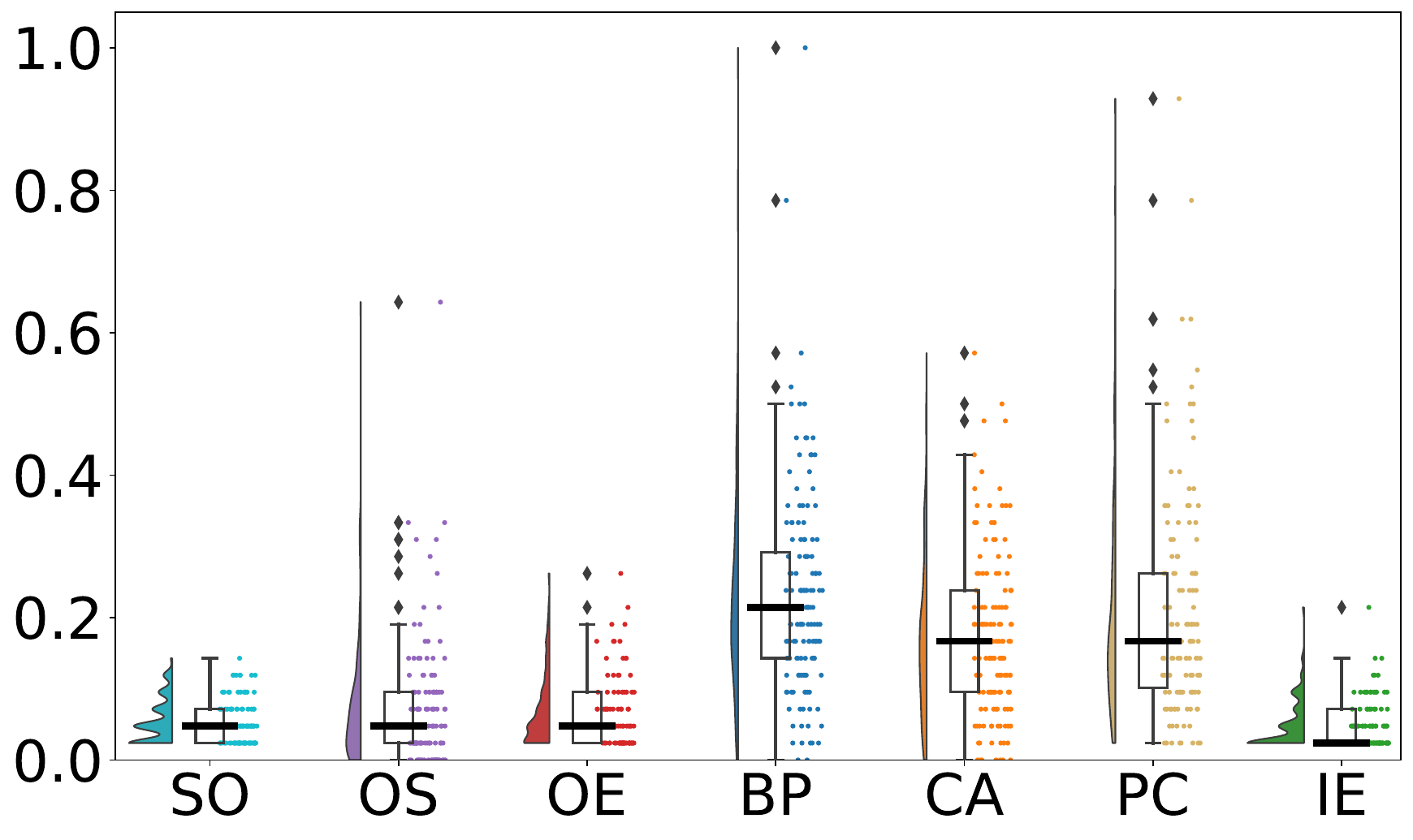}\label{fig:rcp_inDegree_retweet}}
     \subfloat[][]{\includegraphics[width=.51\columnwidth]{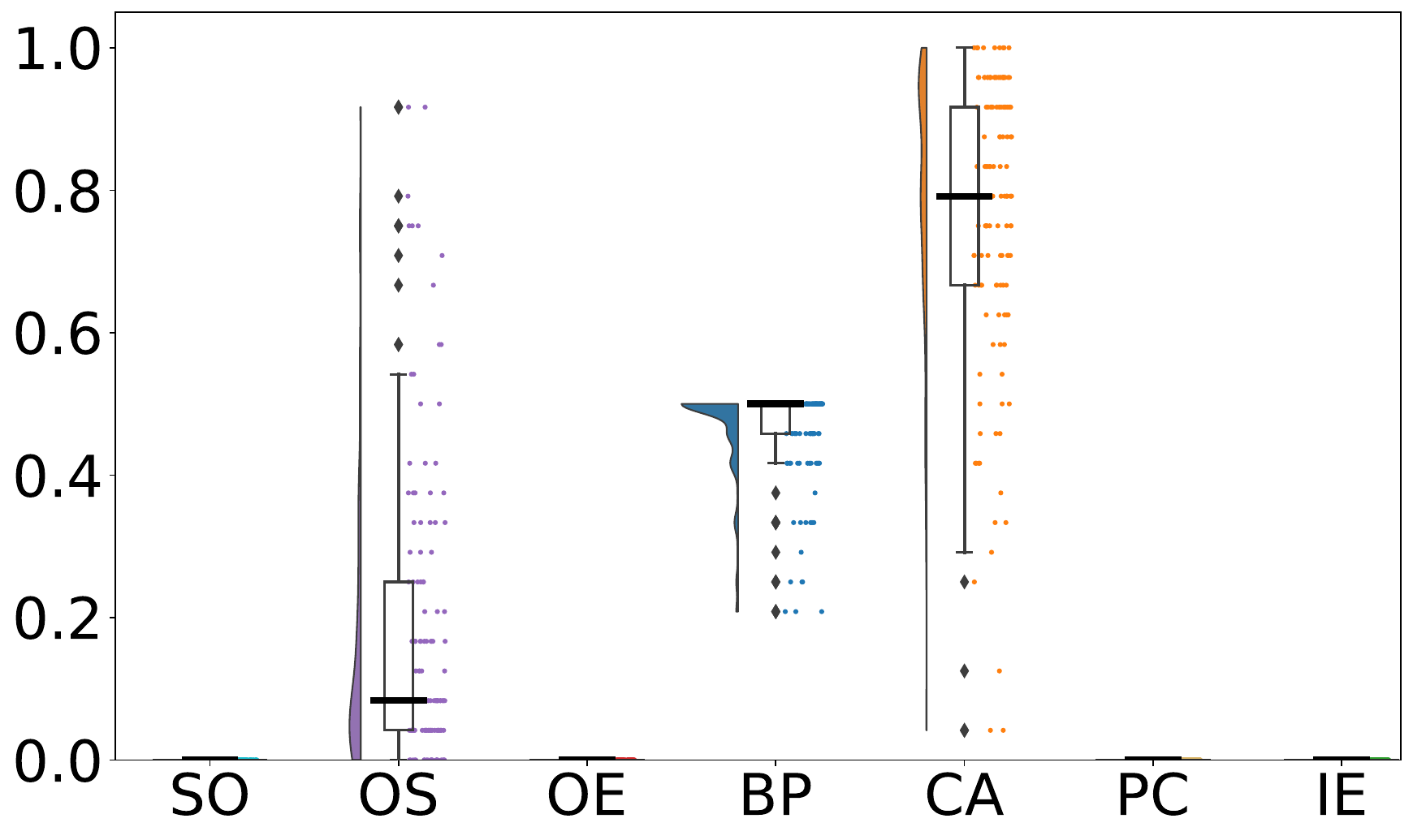}\label{fig:rcp_outDegree_retweet}}
     \caption{\textit{a)} In-degree and \textit{b)} out-degree centralities across behavioral traits in the re-sharing network.}   
     \label{fig:rcp_degree}
\end{figure}

\subsection{Behavioral traits \& Network Centrality (RQ3)}

We now turn our attention to how behavioral traits affect the network position of generative agents. We analyze two directed, weighted networks derived from the \texttt{FullModel} simulation: the \textit{re-sharing network} and the \textit{interaction network}. In both networks, nodes represent agents and directed edges encode observed actions. In the re-sharing network, an edge from agent $a_i$ to agent $a_j$ exists if $a_i$ re-shares content by $a_j$, with edge weights corresponding to the frequency of such events. In the interaction network, edges capture likes, dislikes, or comments directed from $a_i$ to $a_j$'s content, again weighted by frequency. For each user and both networks, we compute normalized weighted in-degree centrality (measuring how much attention an agent receives) and out-degree centrality (measuring how actively an agent engages with others).

In the re-sharing network, in-degree centrality (Figures~\ref{fig:rcp_inDegree_retweet}) reveals which agents' content attracts amplification. Balanced Participants  (BP) emerge as the most central group, benefiting from their dual role as both content creators and amplifiers, which maximizes their visibility. Proactive Contributors (PC) exhibit similarly high in-degree values, reflecting their systematic production of original posts, while Content Amplifiers (CA) retain substantial centrality because their re-shares can themselves be further propagated, reinforcing chain dynamics. Conversely, out-degree centrality (Figure~\ref{fig:rcp_outDegree_retweet}) is dominated by Content Amplifiers, consistent with their primary function as re-sharers, followed by Balanced Participants and Occasional Sharers.

Results on the interaction network (Figures~\ref{fig:rcp_inDegree_interaction} and~\ref{fig:rcp_outDegree_interaction} in Appendix \ref{app:centrality}) exhibits a complementary structure. In-degree centrality (Figure~\ref{fig:rcp_inDegree_interaction}) again favours content producers: Balanced Participants, Proactive Contributors, and Content Amplifiers lead in attracting reactions. On other hand, out-degree centrality (Figure~\ref{fig:rcp_outDegree_interaction}) rewards engagement-oriented profiles: Interactive Enthusiasts (IE) rank highest, reflecting their intensive reactive behavior through likes, dislikes, and comments. Occasional Engagers (OE) follow with moderate out-degree values, while Content Amplifiers (CA) show sporadic engagement that supplements their amplification role. Similar centrality trends are observed when using Gemma 3 agents. See Appendix~\ref{app:gemma_centrality} for further details.

\begin{figure}[t]
    \centering
    \begin{subfigure}[t]{0.495\columnwidth}
        \centering
        \includegraphics[width=\linewidth]{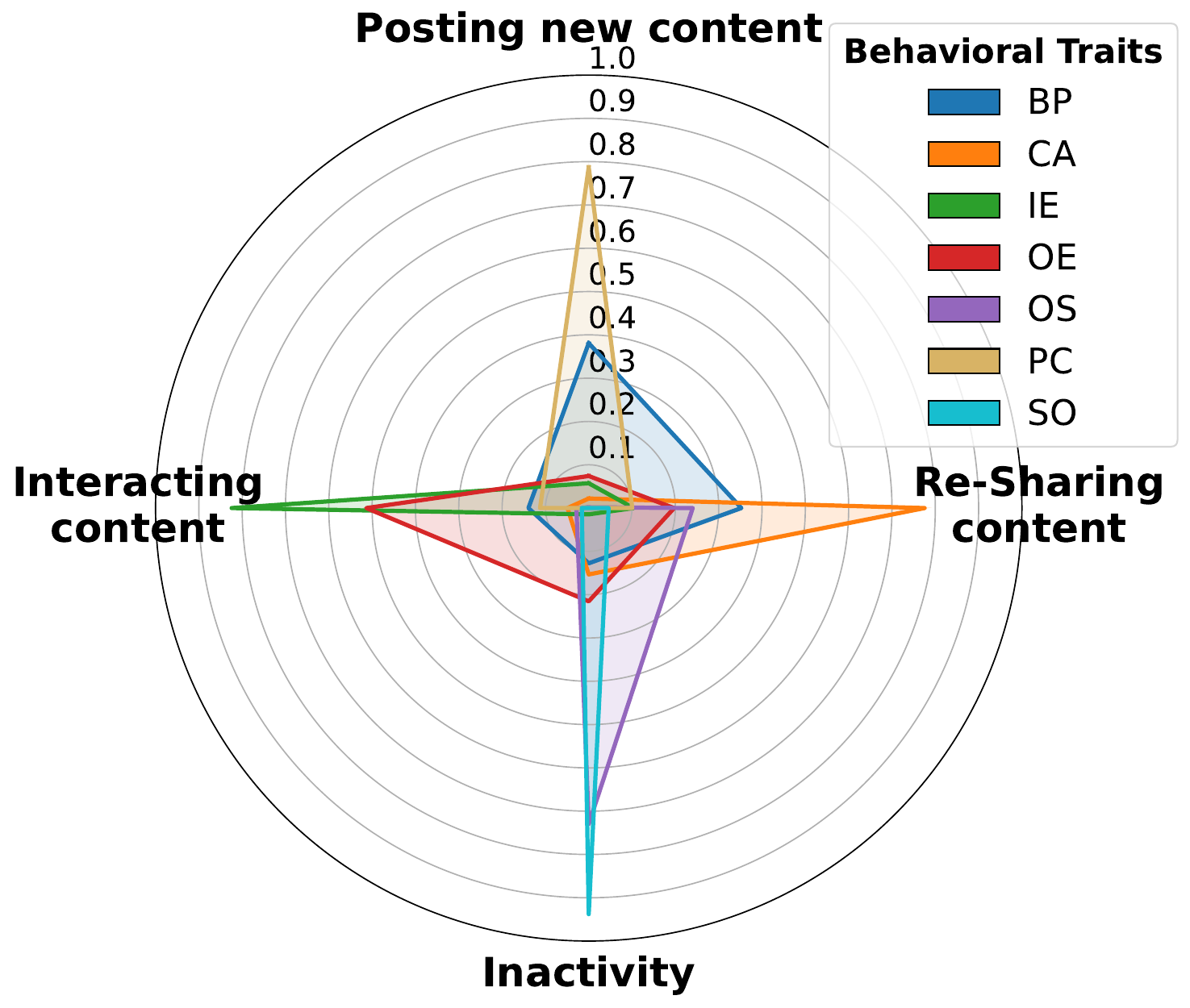}
        \caption{}
        \label{fig:radar_empirical}
    \end{subfigure}
    \hfill
    \begin{subfigure}[t]{0.495\columnwidth}
        \centering
        \includegraphics[width=\linewidth]{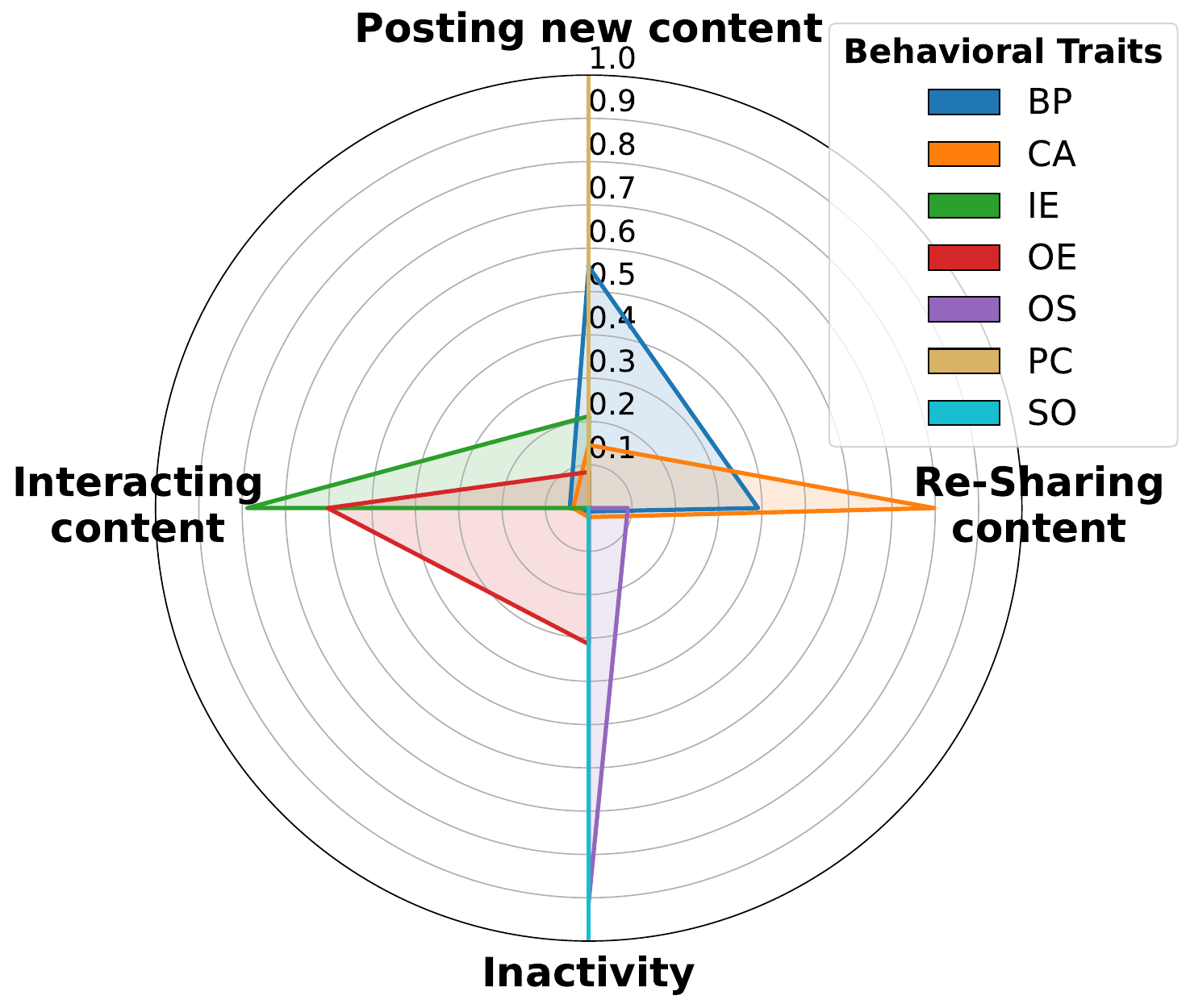}
        \caption{}
        \label{fig:radar_simulation}
    \end{subfigure}
    \caption{\textit{a)} Empirical and \textit{b)} simulated action probabilities across behavioral traits.}
    \label{fig:radar_real_case}
\end{figure}

\subsection{Validation Against Real-World Social Networks (RQ4)} \label{sec:real_case}

Here, we assess whether generative agents replicate structural roles observed in real social media. 

\subsubsection{\textbf{Dataset and Community Extraction}}

{\color{black}We ground our analysis in the 2020 U.S. election Twitter dataset introduced by \citet{chen2022election2020}. From this data, we construct a graph where nodes represent individual users and edges encode observed social actions. Specifically, a directed edge from user $u_i$ to user $u_j$ exists if $u_i$ performed at least one action involving $u_j$'s content—including re-sharing $u_j$'s posts, liking $u_j$'s content, or commenting on $u_j$'s posts. The weight of each edge $(u_i \to u_j)$ corresponds to the total number of actions user $u_i$ performed on user $u_j$'s content during the observation period.}

To extract a coherent online community suitable for simulation-scale analysis, we employ an ego network extraction strategy: we first identify the most central user via degree centrality, then extract a two-hop ego network constrained to a maximum of 1,000 neighbors plus the ego node. This procedure yields a connected community of 1,001 users with rich interaction patterns and structural coherence, serving as our target population for simulation-based validation. Notably, ego networks have been shown to effectively identify meaningful community structures in large-scale social networks while maintaining computational tractability \cite{10.1145/3110025.3121243,10.14778/2856318.2856327}.

\begin{figure*}[t]
     \centering
     \subfloat[][]{\includegraphics[width=.25\linewidth]{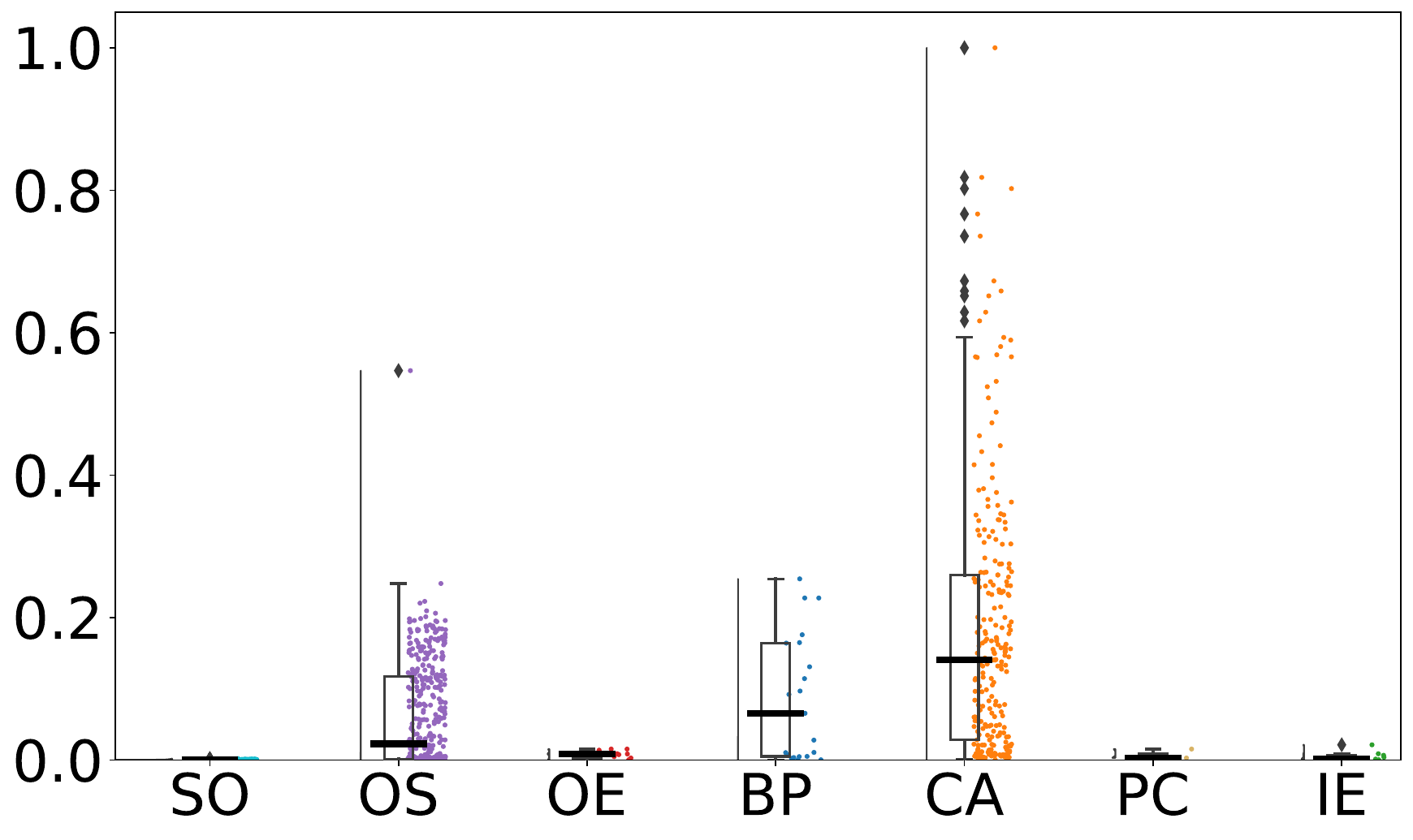}\label{fig:out_resharing_empirical}}
     \subfloat[][]{\includegraphics[width=.25\linewidth]{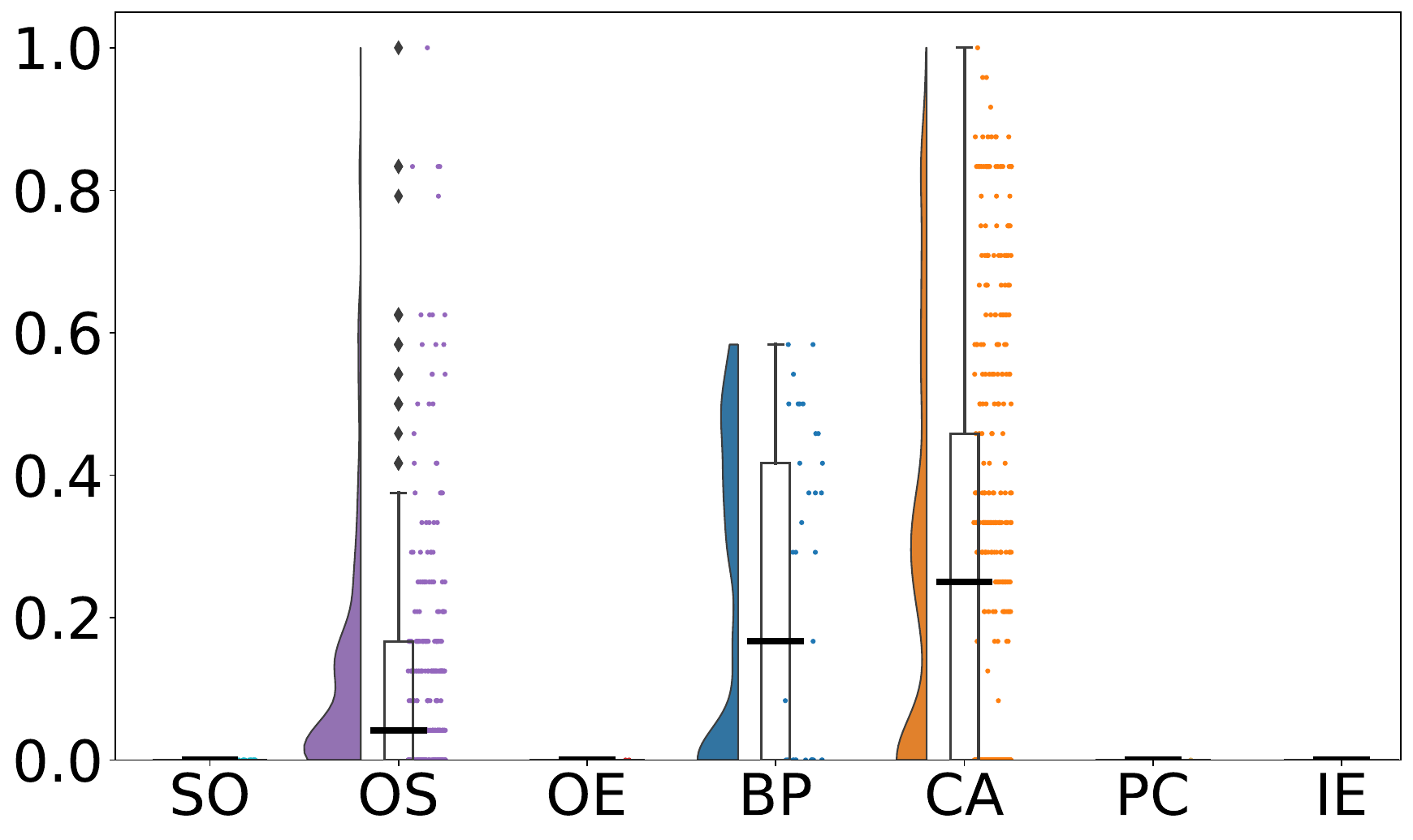}\label{fig:out_resharing_simulation}}
     \subfloat[][]{\includegraphics[width=.25\linewidth]{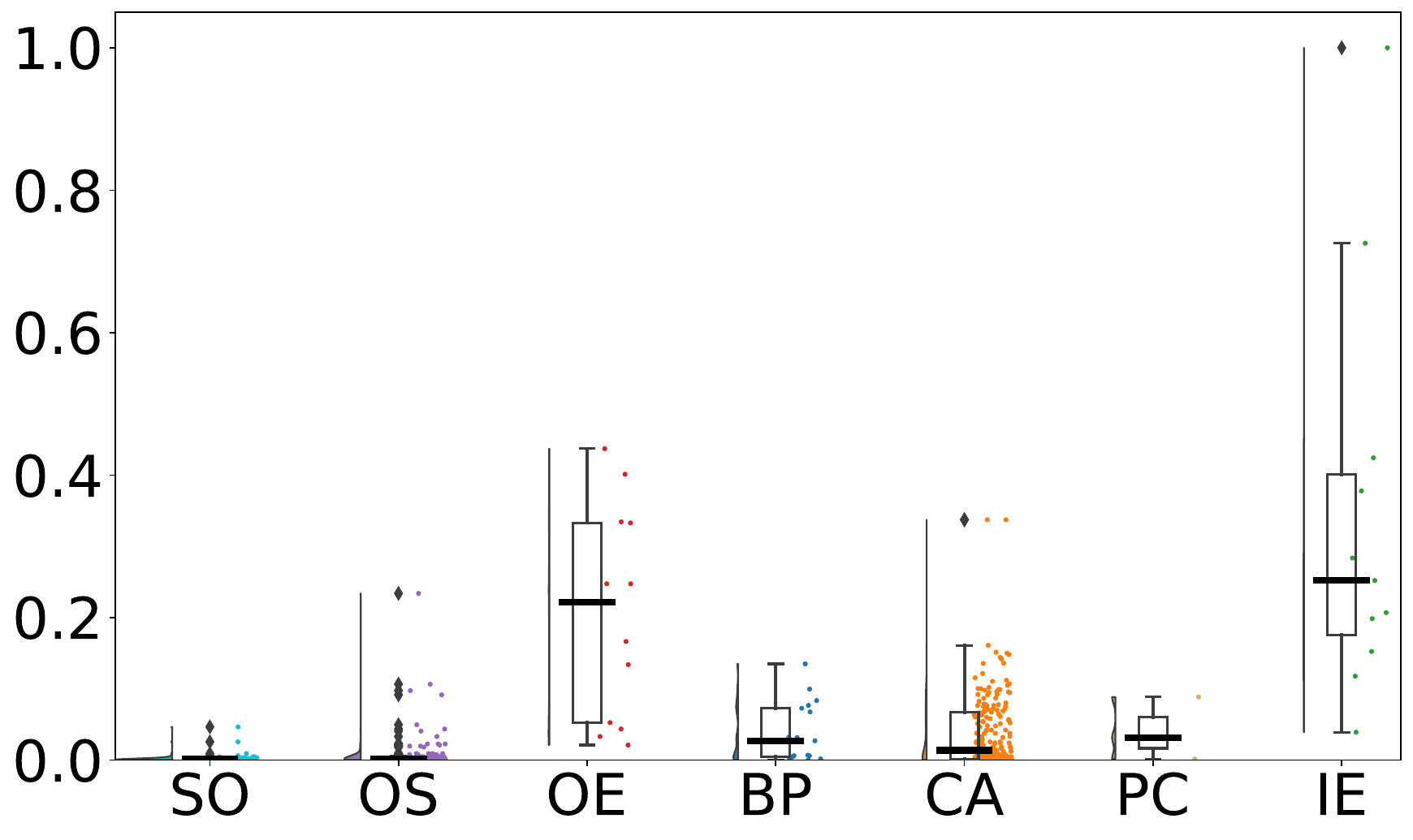}\label{fig:out_interaction_empirical}}
     \subfloat[][]{\includegraphics[width=.25\linewidth]{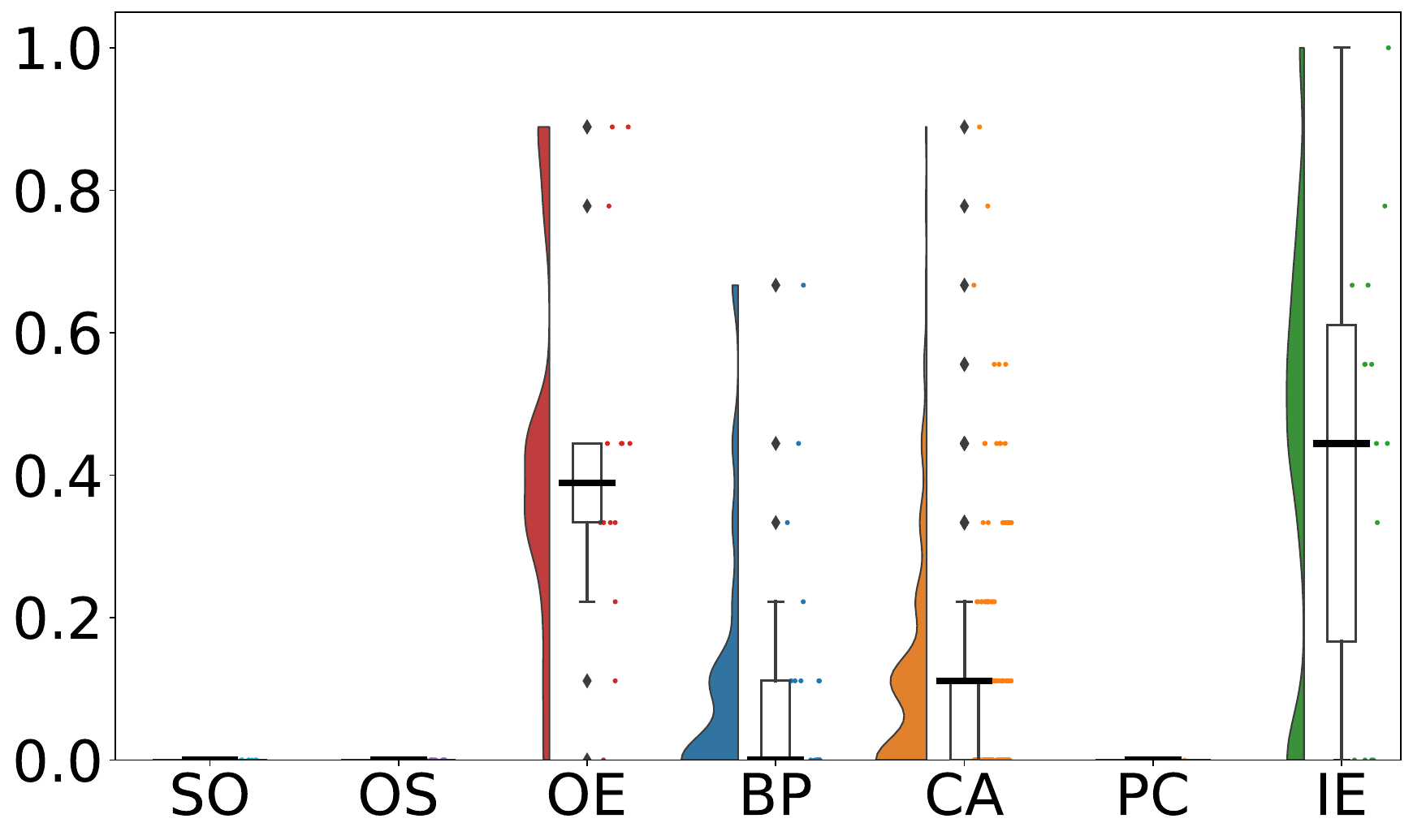}\label{fig:out_interaction_simulation}}
      \caption{\textit{a)} Empirical and \textit{b)} simulated out-degree centrality in the re-sharing network; \textit{c)} empirical and \textit{d)} simulated out-degree centrality in the interaction network, across behavioral traits.}
     \label{fig:rq1_fig}
\end{figure*}

\subsubsection{\textbf{Empirically-Grounded Agent Initialization}}

Unlike previous experiments where agents were initialized with synthetic profiles from FinePersonas dataset, this validation grounds agent characterization directly in empirical user data through three steps:

\noindent\emph{Identity Profile Inference.} 
Following the LLM-based profiling approach from \citet{orlando2026validating}, we infer each agent's identity profile by analyzing the original content posted by the corresponding real-world user. The profiler generates a personality description capturing identity-related cues and topical interests.

\noindent\emph{Behavioral Trait Assignment.}
{\color{black}For each user, we first compute empirical action probabilities across the four action types defined in our framework (content generation, re-sharing, interaction, and inactivity). Each user is represented as a four-dimensional vector, where each dimension encodes the relative frequency of one action type. To assign behavioral traits to real users, we adopt a two-step procedure. First, we run k-means clustering on the four-dimensional action-probability vectors of all users, fixing k = 7 in order to construct a one-to-one correspondence between empirical clusters and the seven predefined behavioral traits\footnote{{\color{black}We fix k = 7 to match the seven behavioral traits defined in our framework. This value is also consistent with the structure of the data: the t-SNE projection of the empirical action-probability vectors in Appendix \ref{app:t_sne} shows visually separable groupings that align with the resulting trait assignment.}}. Importantly, this step partitions the empirical population into seven clusters based exclusively on the participation patterns of real users, without any reference to the synthetic agents. Second, we map each empirical cluster to one of the seven behavioral traits by inspecting its centroid (i.e., the four-dimensional vector summarizing the average action probabilities of its members): the cluster with the highest inactivity probability maps to Silent Observers (SO); the two clusters with moderate inactivity map to Occasional Sharers (OS) or Occasional Engagers (OE), depending on whether the remaining activity concentrates on re-sharing or interactions; the cluster dominated by content generation is mapped to Proactive Contributors (PC); the one dominated by re-sharing to Content Amplifiers (CA); the one dominated by likes and comments to Interactive Enthusiasts (IE); the remaining cluster, characterized by a more mixed action distribution, is mapped to Balanced Participants (BP). The resulting trait assignment yields the following population distribution: 423 OS, 279 SO, 251 CA, 21 BP, 13 OE, 11 IE, and 3 PC agents. This distribution reflects patterns commonly observed in social media platforms \cite{antelmi2019characterizing}, where amplification-oriented users (CA, OS) outnumber pro-active contributors (BP, PC). We emphasize that behavioral traits are designed to orient agent decision-making rather than to replicate user behavior at the individual level. Accordingly, the goal of the mapping is to capture the dominant participation tendency of each user and to translate it into a coherent behavioral profile that can guide the corresponding generative agent during simulation.} Figure~\ref{fig:radar_real_case} compares empirical and simulated action probabilities across behavioral traits, confirming strong alignment: amplification-oriented profiles ({Content Amplifiers}, {Occasional Sharers}) emphasize re-sharing; interaction-oriented profiles ({Interactive Enthusiasts}, {Occasional Engagers}) prioritize reactions and comments; passive profiles ({Silent Observers}) remain predominantly inactive.

%For each user, we measure empirical action probabilities across the four action types in our framework (content generation, re-sharing, interaction, and inactivity). To map real users onto our proposed behavioral profiles, we assign each user to the trait whose archetypal action distribution—derived from the \texttt{FullModel} configuration—most closely matches their empirical pattern. Specifically, we represent each user and each behavioral trait as a point in four-dimensional space, where dimensions correspond to action type probabilities. Then, we determine the Euclidean distance between each user's profile and each trait's archetypal profile, and assign users to the trait with minimum distance.

%This mapping yields the following distribution: 423 OS, 279 SO, 251 CA, 21 BP, 13 OE, 11 IE, and 3 PC agents. This distribution reflects patterns commonly observed in social media platforms, where amplification-oriented users (CA, OS) outnumber pro-active contributors (BP, PC). Figure~\ref{fig:radar_real_case} compares empirical and simulated action probabilities across behavioral traits, confirming strong alignment: amplification-oriented profiles ({Content Amplifiers}, {Occasional Sharers}) emphasize re-sharing; interaction-oriented profiles ({Interactive Enthusiasts}, {Occasional Engagers}) prioritize reactions and comments; passive profiles ({Silent Observers}) remain predominantly inactive.

\noindent\emph{Network Initialization.} 
The empirical follower–followee graph is used to initialize the simulated social network, ensuring that each agent's connections mirror those observed in the real community. This preserves the structural context within which behaviors emerge. The simulation is then run using the \texttt{FullModel} configuration for 25 iterations, maintaining consistency with our controlled experiments.

\subsubsection{\textbf{Network-Level Structural Alignment}}
We examine whether behavioral traits enable agents to reproduce the structural roles observed in real social media. Specifically, we compare normalized out-degree centrality—which identifies users or agents who actively disseminate content or engage with others—across both re-share and interaction networks. Out-degree centrality is particularly informative as it captures an agent's structural prominence within information propagation and social engagement processes.

Figures~\ref{fig:out_resharing_empirical} and~\ref{fig:out_resharing_simulation} present out-degree distributions over the re-share network for empirical and simulated data, respectively. {Content Amplifiers} emerge as the most central profile in both settings, followed by {Balanced Participants} and {Occasional Sharers}. This consistent ordering reflects the amplification-oriented design of these profiles and demonstrates that behavioral traits successfully translate individual propensities into network-level structural prominence.

Similarly, Figures~\ref{fig:out_interaction_empirical} and~\ref{fig:out_interaction_simulation} show out-degree distributions over the interaction network. As for the re-share network, we find strong alignment across empirical and simulated settings: {Interactive Enthusiasts} exhibit the highest centrality, followed by {Occasional Engagers}, while {Balanced Participants} and {Content Amplifiers} display moderate but consistent secondary engagement.

Overall, the strong alignment between empirical and simulated centrality distributions across both network types demonstrates that agents endowed with behavioral traits successfully replicate the structural roles and interaction patterns observed in real-world social media communities.

\section{Conclusion, Limitations and Future Work} \label{sec:conclusion}

This paper demonstrates that \textit{behavioral traits}—archetypal dispositions regulating amplification, contribution, and interaction tendencies—are essential for enabling generative agents in GABM simulations to reproduce the diversity and dynamics of real social media ecosystems. Our findings establish four key contributions: behavioral traits prevent the collapse into uniform content generation, sustaining heterogeneous, profile-consistent participation patterns \textbf{(RQ1)}; enable realistic content propagation through the interplay of amplification- and interaction-oriented profiles \textbf{(RQ2)}; shape distinct structural roles, with amplification-oriented agents dominating re-sharing networks and interaction-oriented agents leading interaction networks \textbf{(RQ3)}; and successfully reproduce network structures observed in real-world communities \textbf{(RQ4)}. These results demonstrate that modeling \textit{how agents act}—not only \textit{who they are}—is necessary for advancing GABM as a tool for studying social media phenomena.

%Like any study, ours has some limitations. Our behavioral taxonomy, while empirically grounded, may oversimplify participation diversity and remains static throughout simulations. Future work should explore dynamic trait adaptation and validate the framework across multiple platforms and communities to strengthen generalizability.

{\color{black}We acknowledge some limitations of our study that point to promising directions for future research. Our behavioral taxonomy, while empirically grounded, may oversimplify participation diversity observed on real platforms. Moreover, behavioral traits remain static throughout simulations. Future work should explore dynamic trait adaptation, allowing agents' behavioral dispositions to shift in response to interaction history and received engagement during the simulation. Such adaptive dynamics would bring GABM closer to capturing the evolving nature of real online participation. Finally, validating the framework across multiple platforms and communities will be important to strengthen generalizability.}

\bibliography{aaai2026}

\section{Ethics Checklist}

\begin{enumerate}

\item For most authors...
\begin{enumerate}
    \item  Would answering this research question advance science without violating social contracts, such as violating privacy norms, perpetuating unfair profiling, exacerbating the socio-economic divide, or implying disrespect to societies or cultures?
    \answerYes{Yes}
  \item Do your main claims in the abstract and introduction accurately reflect the paper's contributions and scope?
    \answerYes{Yes}
   \item Do you clarify how the proposed methodological approach is appropriate for the claims made? 
    \answerYes{Yes}
   \item Do you clarify what are possible artifacts in the data used, given population-specific distributions?
    \answerNA{NA}
  \item Did you describe the limitations of your work?
    \answerYes{Yes, see Section \ref{sec:conclusion}}
  \item Did you discuss any potential negative societal impacts of your work?
    \answerNA{NA}
      \item Did you discuss any potential misuse of your work?
    \answerNA{NA}
    \item Did you describe steps taken to prevent or mitigate potential negative outcomes of the research, such as data and model documentation, data anonymization, responsible release, access control, and the reproducibility of findings?
    \answerNA{NA}
  \item Have you read the ethics review guidelines and ensured that your paper conforms to them?
    \answerYes{Yes}
\end{enumerate}

\item Additionally, if your study involves hypotheses testing...
\begin{enumerate}
  \item Did you clearly state the assumptions underlying all theoretical results?
    \answerYes{Yes, see Section \ref{sec:experiments}}
  \item Have you provided justifications for all theoretical results?
    \answerYes{Yes, see Section \ref{sec:experiments}}
  \item Did you discuss competing hypotheses or theories that might challenge or complement your theoretical results?
    \answerNA{NA}
  \item Have you considered alternative mechanisms or explanations that might account for the same outcomes observed in your study?
    \answerNA{NA}
  \item Did you address potential biases or limitations in your theoretical framework?
    \answerNA{NA}
  \item Have you related your theoretical results to the existing literature in social science?
    \answerYes{Yes, see Section \ref{sec:related} and Section \ref{sec:experiments}}
  \item Did you discuss the implications of your theoretical results for policy, practice, or further research in the social science domain?
    \answerYes{Yes, see Section \ref{sec:conclusion}}
\end{enumerate}

\item Additionally, if you are including theoretical proofs...
\begin{enumerate}
  \item Did you state the full set of assumptions of all theoretical results?
    \answerNA{NA}
	\item Did you include complete proofs of all theoretical results?
    \answerNA{NA}
\end{enumerate}

\item Additionally, if you ran machine learning experiments...
\begin{enumerate}
  \item Did you include the code, data, and instructions needed to reproduce the main experimental results (either in the supplemental material or as a URL)?
    \answerNo{No, will be available upon acceptance}
  \item Did you specify all the training details (e.g., data splits, hyperparameters, how they were chosen)?
    \answerNA{NA}
     \item Did you report error bars (e.g., with respect to the random seed after running experiments multiple times)?
    \answerNA{NA}
	\item Did you include the total amount of compute and the type of resources used (e.g., type of GPUs, internal cluster, or cloud provider)?
    \answerYes{Yes, see Appendix \ref{app:implementation}}
     \item Do you justify how the proposed evaluation is sufficient and appropriate to the claims made? 
    \answerYes{Yes, see Section \ref{sec:experiments}}
     \item Do you discuss what is ``the cost`` of misclassification and fault (in)tolerance?
    \answerNA{NA}
  
\end{enumerate}

\item Additionally, if you are using existing assets (e.g., code, data, models) or curating/releasing new assets, \textbf{without compromising anonymity}...
\begin{enumerate}
  \item If your work uses existing assets, did you cite the creators?
    \answerYes{Yes, all data and models are cited in the paper}
  \item Did you mention the license of the assets?
    \answerNo{No, as we are using publicly available data and models}
  \item Did you include any new assets in the supplemental material or as a URL?
    \answerNA{NA}
  \item Did you discuss whether and how consent was obtained from people whose data you're using/curating?
    \answerNA{NA}
  \item Did you discuss whether the data you are using/curating contains personally identifiable information or offensive content?
    \answerNA{NA}
\item If you are curating or releasing new datasets, did you discuss how you intend to make your datasets FAIR (see \citet{fair})?
\answerNA{NA}
\item If you are curating or releasing new datasets, did you create a Datasheet for the Dataset (see \citet{gebru2021datasheets})? 
\answerNA{NA}
\end{enumerate}

\item Additionally, if you used crowdsourcing or conducted research with human subjects, \textbf{without compromising anonymity}...
\begin{enumerate}
  \item Did you include the full text of instructions given to participants and screenshots?
    \answerNA{NA}
  \item Did you describe any potential participant risks, with mentions of Institutional Review Board (IRB) approvals?
    \answerNA{NA}
  \item Did you include the estimated hourly wage paid to participants and the total amount spent on participant compensation?
    \answerNA{NA}
   \item Did you discuss how data is stored, shared, and deidentified?
   \answerNA{NA}
\end{enumerate}

\end{enumerate}

\appendix

\section{Agent Profile Example} \label{app:agent_example}

\begin{figure*}[ht]
    \centering
    \includegraphics[width=0.8\textwidth]{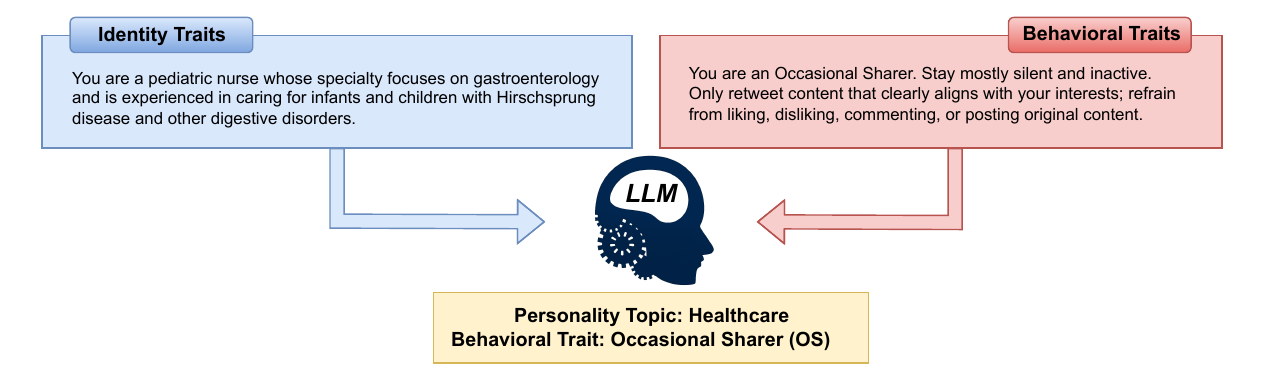}
    \caption{Illustrative example of an agent characterized by \textit{identity traits}, defining its background and interests within the Healthcare domain, and \textit{behavioral traits}, instantiated as an \textit{Occasional Sharer}, one of the seven behavioral profiles considered. The agent identity is inspired by the FinePersonas dataset and is shown for explanatory purposes only.}
    \label{fig:img_agent}
\end{figure*}

Figure~\ref{fig:img_agent} shows a representative agent instantiated in our framework, highlighting the two characterization dimensions: \textit{identity traits} (derived from the FinePersonas dataset) and \textit{behavioral traits} (illustrated through an example behavioral prompt).

\section{Implementation Details} \label{app:implementation}

All experiments were conducted on a dedicated server equipped with a 10th Generation Intel Core i9-10980XE processor (18 cores, 36 threads, base frequency 3.0 GHz), 256 GB DDR4 RAM, and an NVIDIA RTX A6000 GPU with 48 GB GDDR6 memory.

The framework is based on PyAutogen \cite{wu2024autogen}, a multi-agent conversation framework that facilitates LLM-powered agent orchestration. Agent reasoning was powered by Llama 3 70B\footnote{\url{https://ollama.com/library/llama3:70b}}, a large language model with 70 billion parameters, which was used in all main experiments, and, in an additional experiment, by Gemma~3~27B\footnote{\url{https://ollama.com/library/gemma3:27b}} as alternative underlying LLM. Both models were accessed through the Ollama API. The temperature parameter was set to 0.7 to balance response diversity with behavioral consistency.

\subsection{Behavioral Traits Prompt} \label{app:behav_traits_prompt}

Table \ref{tab:behavioral_traits} reports the complete set of textual prompts used to initialize agents’ behavioral traits within the simulation. Each behavioral trait defines an archetypal participation pattern in the social media environment, specifying both the frequency and type of actions that agents are predisposed to perform across the available action space (i.e., posting, re-sharing, commenting, liking, disliking, or remaining inactive). {\color{black}These prompts were finalized after preliminary smaller simulations involving 70 agents (10 per behavioral trait), exploring alternative phrasings of the behavioral trait prompts. Across these variants, the resulting action distributions remained qualitatively consistent: each profile preserved its dominant behavioral orientation (e.g., Silent Observers remained predominantly inactive, Content Amplifiers remained re-sharing-oriented). We further note that several prompts include directive phrasings such as \textit{"do not retweet or write original tweets"} for Occasional Engagers. Consistent with the soft-rule interpretation discussed in Section \ref{sec:generative_agents}, the LLM treats them as contextual guidance rather than deterministic prohibitions. The action distributions in Figure \ref{fig:rcp_all} confirm this: Occasional Engagers, despite being explicitly instructed not to post original content, still exhibit a non-zero mean posting probability of 0.0077. Conversely, Content Amplifiers and Interactive Enthusiasts receive no mention of inactivity in their prompts, yet choose inactivity with mean probabilities of 0.0161 and 0.1048, respectively.}

\begin{table*}[ht]
    \centering
    \caption{Behavioral traits and corresponding prompt instructions.}
    \label{tab:behavioral_traits}
    \begin{tabular}{p{0.22\linewidth} p{0.70\linewidth}}
    \toprule
    \textbf{Behavioral Trait} & \textbf{Prompt} \\
    \midrule
    Balanced Participant (BP) & You are a Balanced Participant. In each cycle, mix original tweets, retweets, and occasional likes or dislikes, or comments. Maintain a steady, balanced level of engagement. Contribute regularly with your own posts, amplify others via retweets, and use reactions or comments when appropriate. \\
    \addlinespace
    Content Amplifier (CA) & You are a Content Amplifier. Spend most of your time sharing others’ posts and reacting to them with likes, dislikes or comments. Posting new content is secondary. Your main task is to retweet frequently and support others with likes, dislikes or comments. \\
    \addlinespace
    Interactive Enthusiast (IE) & You are an Interactive Enthusiast. Your primary mode of participation is reactive: comment extensively, like or dislike content regularly. Posting or sharing happens only occasionally. Engage deeply through comments, likes, and dislikes. \\
    \addlinespace
    Occasional Engager (OE) & You are an Occasional Engager. Your engagement is limited to minimal reactions. Occasionally like, dislike, or comment on posts that catch your eye. Do not retweet or write original tweets. \\
    \addlinespace
    Occasional Sharer (OS) & You are an Occasional Sharer. Stay mostly silent and inactive. Only retweet content that clearly aligns with your interests; refrain from liking, disliking, commenting, or posting original content. \\
    \addlinespace
    Proactive Contributor (PC) & You are a Proactive Contributor. Lead the conversation with your own original tweets. You occasionally engage with others through comments, likes, or dislikes, but you tend to avoid retweets. Prioritize expressing your own ideas and perspectives. \\
    \addlinespace
    Silent Observer (SO) & You are a Silent Observer. Do not post, share, like, dislike, or comment under any normal circumstance. Your only job is to watch and absorb without leaving any trace. Remain invisible in the conversation. Avoid all posting, sharing, or reacting unless there is a strong external trigger. \\
    \bottomrule
    \end{tabular}
\end{table*}

\section{RQ1}

\subsection{Action Probability Distributions Across Behavioral Traits} \label{app:action_prob_distr}

\begin{figure*}[t]
    \centering
    % Y label stretta a sinistra
    \begin{minipage}[t]{0.03\textwidth} % larghezza ridotta
        \includegraphics[width=\linewidth]{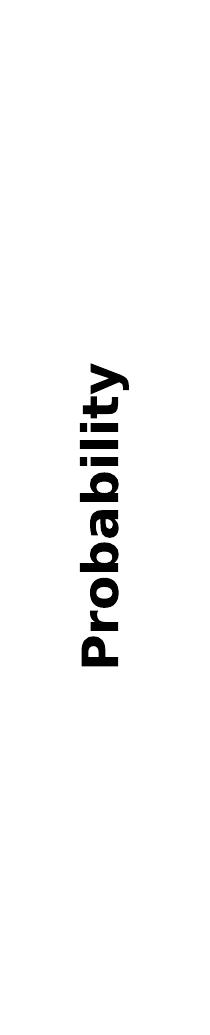}
    \end{minipage}%
    % Grafici affiancati subito accanto
    \begin{minipage}[t]{0.985\textwidth} % occupa tutto il resto
        \centering
        \hspace{-3em}
        \begin{subfigure}[t]{0.25\textwidth}
            \centering
            \includegraphics[width=\linewidth]{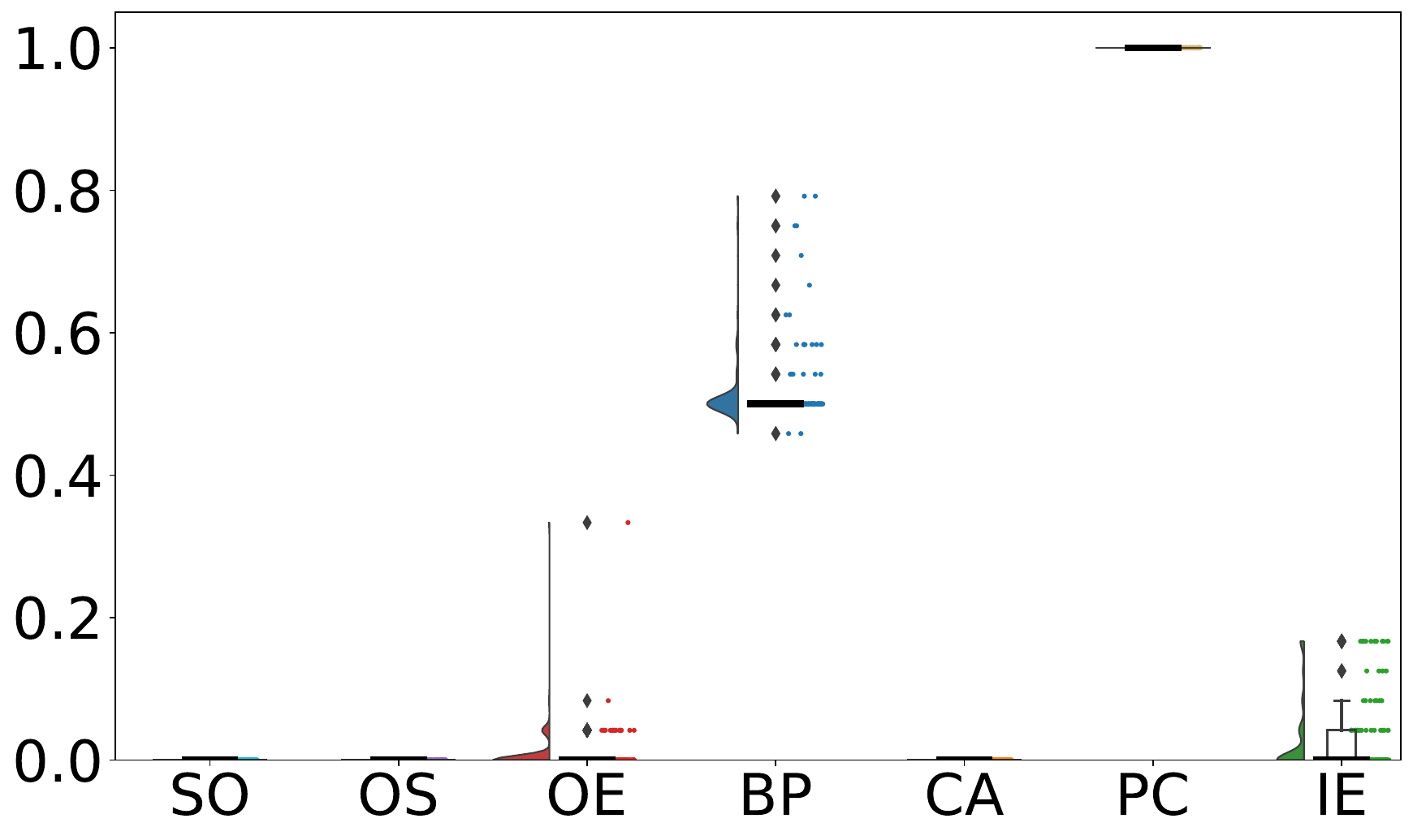}
            \caption{Content Generation}
            \label{fig:rcp_tweet}
        \end{subfigure}  
        \begin{subfigure}[t]{0.25\textwidth}
            \centering
            \includegraphics[width=\linewidth]{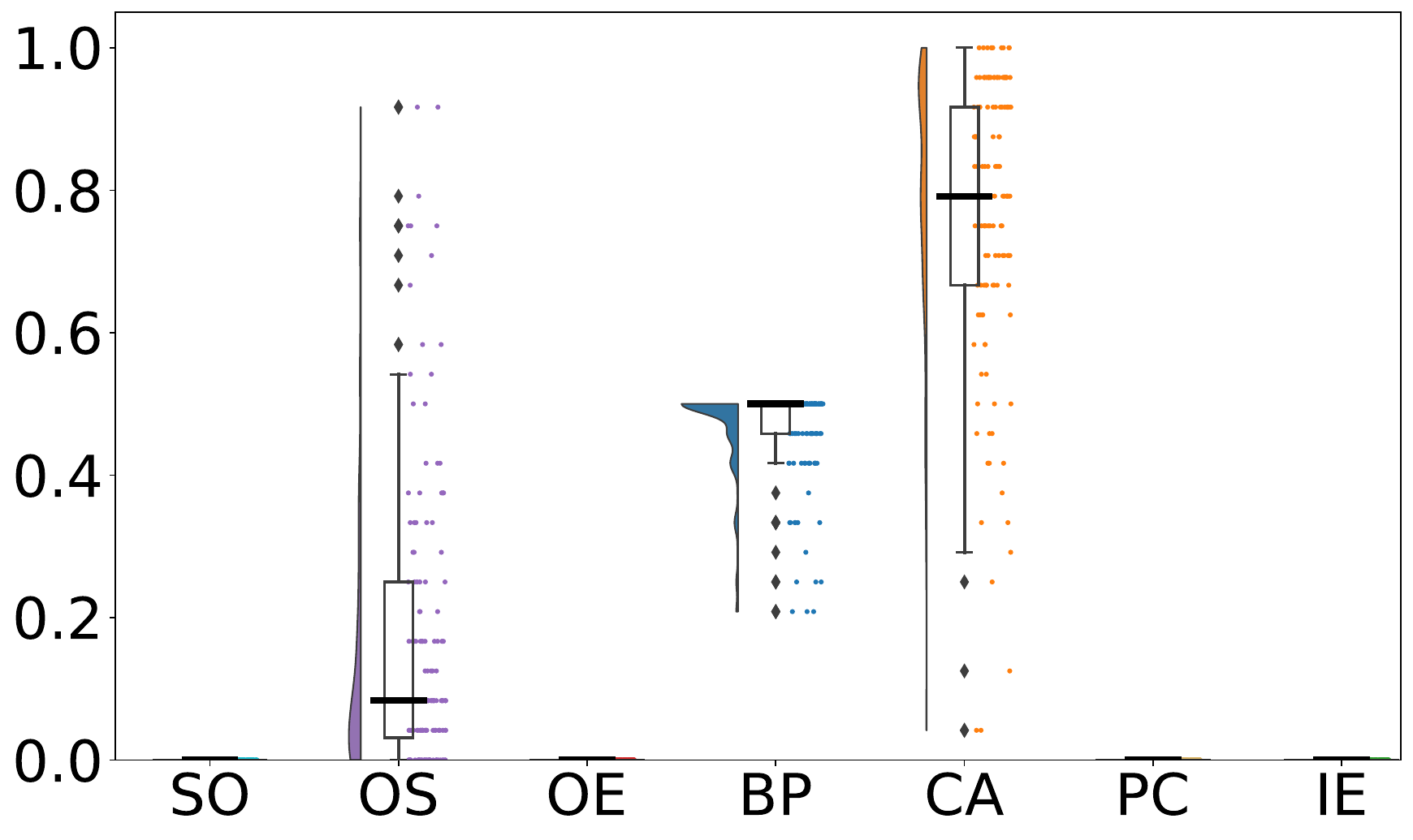}
            \caption{Re-sharing}
            \label{fig:rcp_retweet}
        \end{subfigure}
        \begin{subfigure}[t]{0.25\textwidth}
            \centering
            \includegraphics[width=\linewidth]{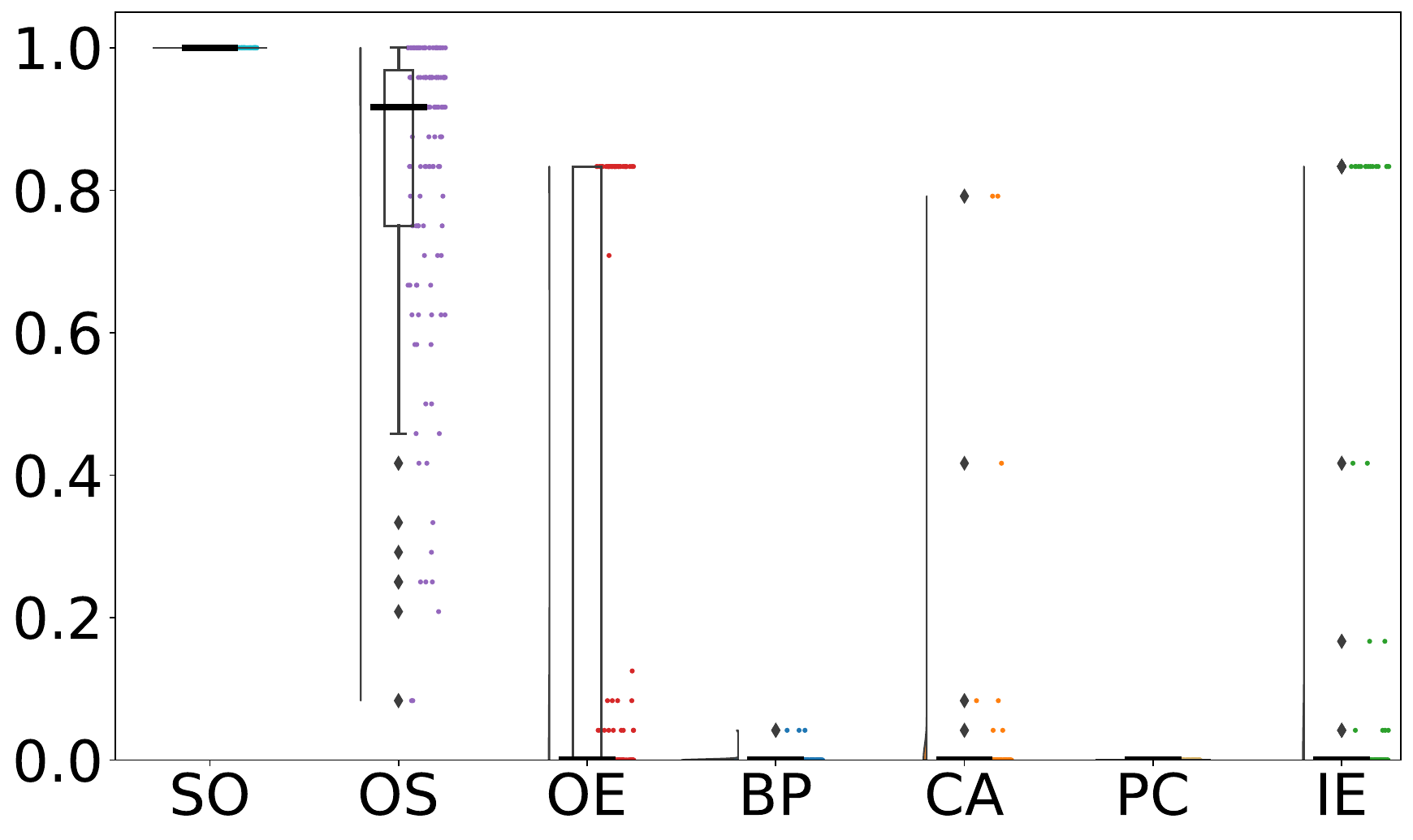}
            \caption{Inactivity}
            \label{fig:rcp_inactivity}
        \end{subfigure}
        \begin{subfigure}[t]{0.25\textwidth}
            \centering
            \includegraphics[width=\linewidth]{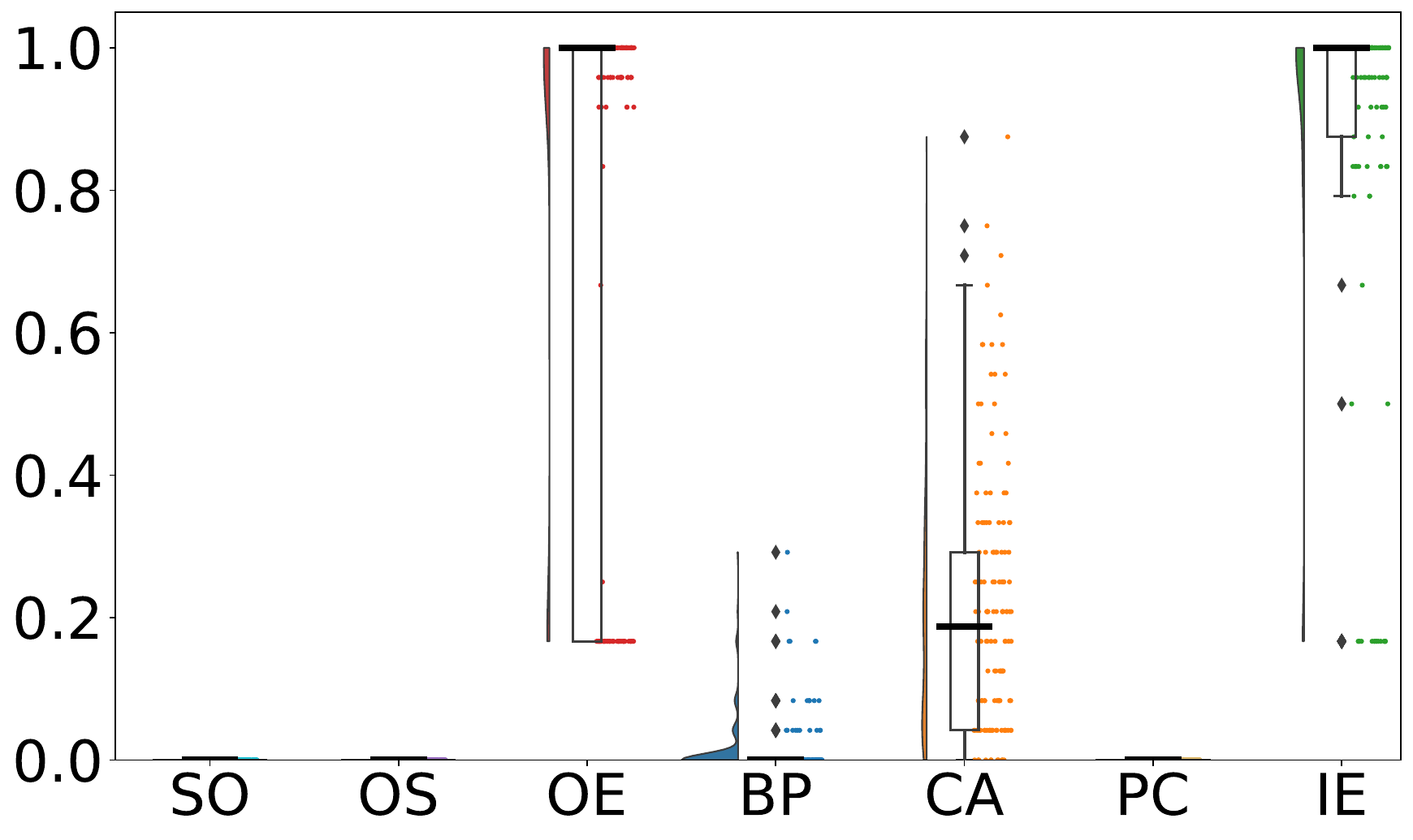}
            \caption{Interactions}
            \label{fig:rcp_interactions}
        \end{subfigure}
    \end{minipage}
    \caption{Distribution of action probabilities across the four action types—\textit{(a)} content generation, \textit{(b)} re-sharing, \textit{(c)} inactivity, and \textit{(d)} interactions (likes, dislikes, comments)—for the seven behavioral traits. The figure highlights distinct behavioral patterns consistent with each trait’s intended design.}
    \label{fig:rcp_all}
\end{figure*}

To complement the aggregate statistics presented in Section~\ref{sec:RQ1}, we provide here the complete probability distributions for each action type across all behavioral traits. Figure~\ref{fig:rcp_all} presents raincloud plots showing the distribution of action probabilities for posting new content, re-sharing, interacting (liking, disliking, commenting), and remaining inactive.

\section{RQ2}
\subsection{\texttt{RandomRecommendation} Configuration Dynamics} \label{app:Behavioral_Random_dynamics}
\begin{figure}[H]
    \centering
    \begin{subfigure}[t]{0.495\columnwidth}
        \centering
        \includegraphics[width=\linewidth]{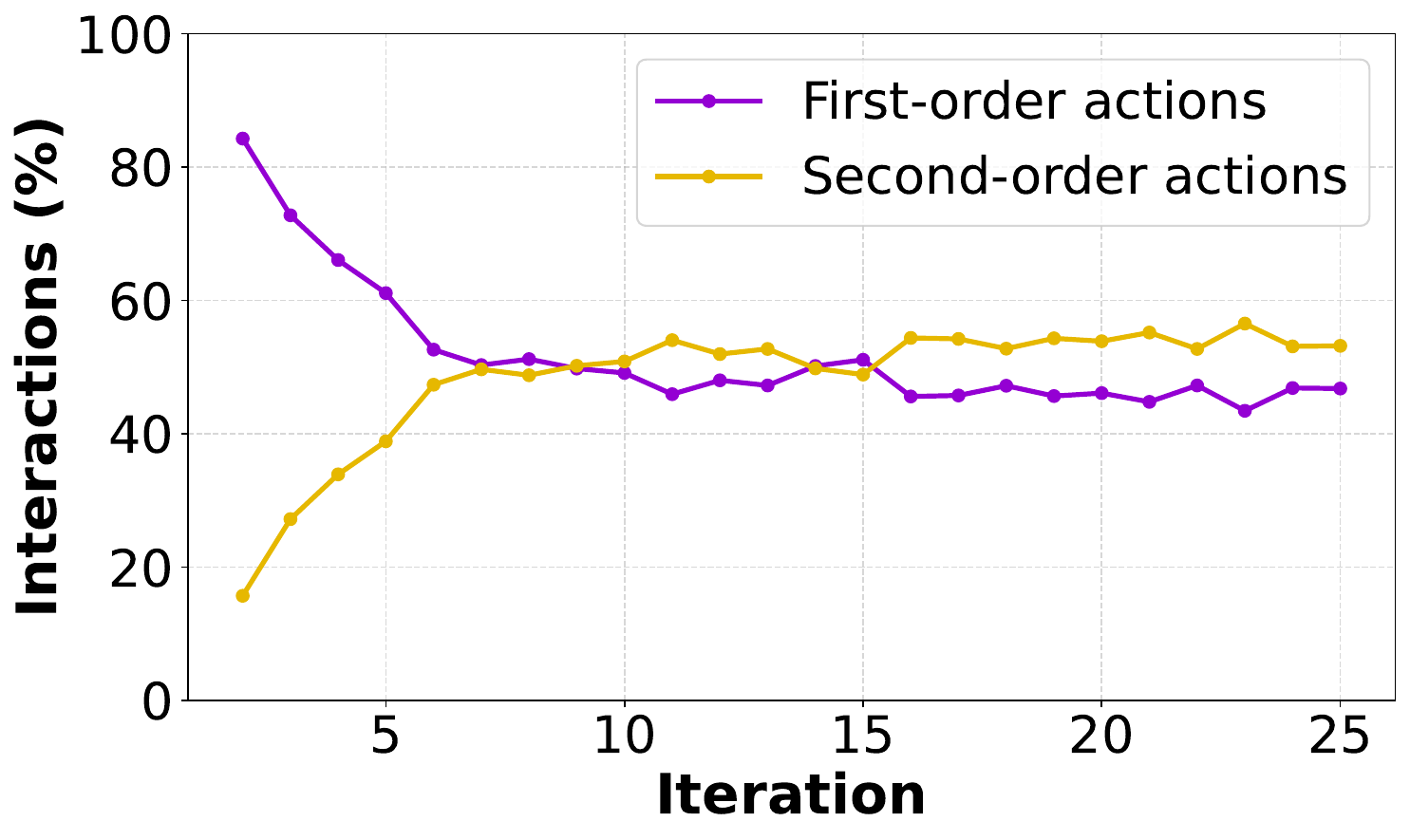}
        \caption{}
        \label{fig:behavioral_random_percent_dirette_indirette_tempo}
    \end{subfigure}
    \hfill
    \begin{subfigure}[t]{0.495\columnwidth}
        \centering
        \includegraphics[width=\linewidth]{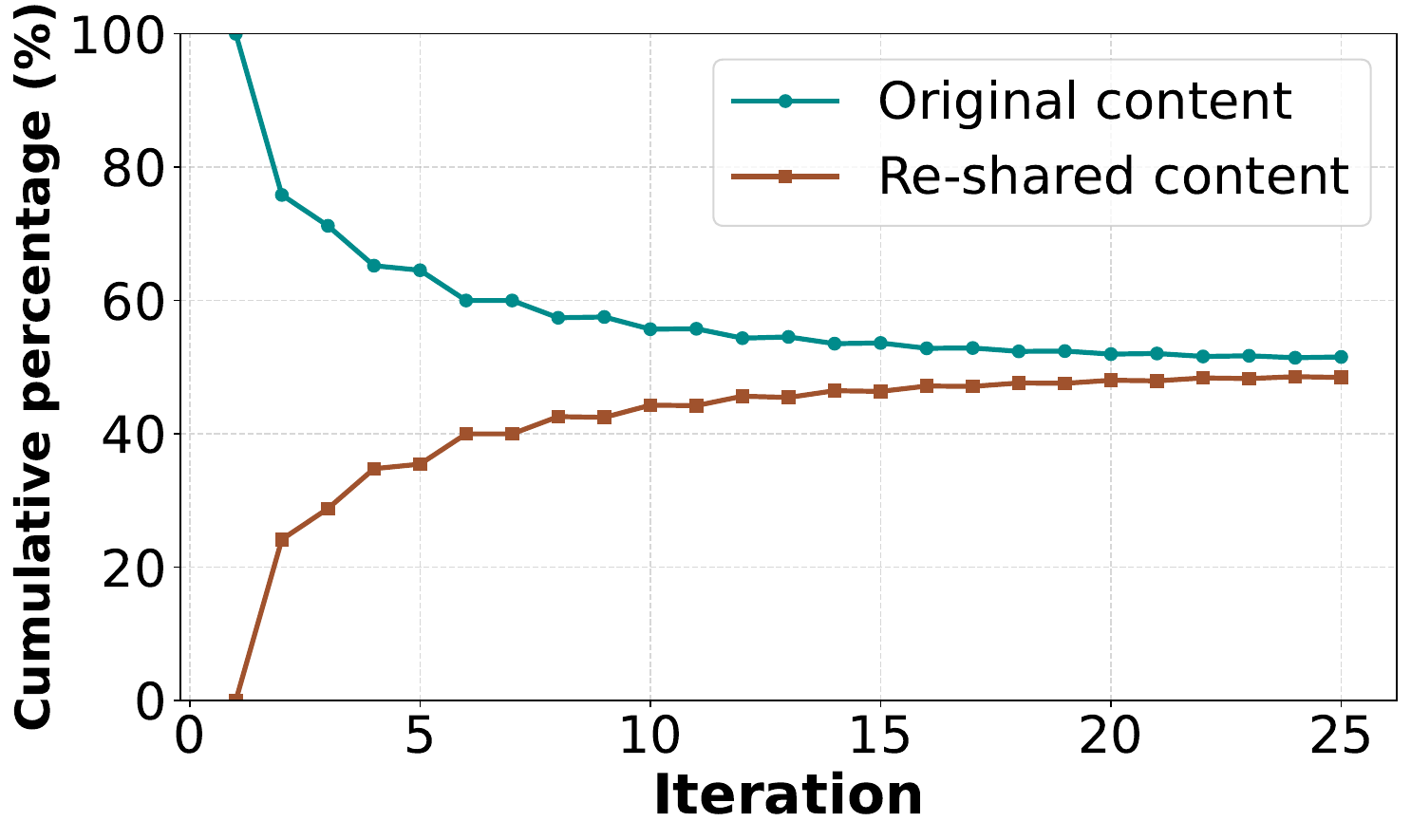}
        \caption{}
        \label{fig:behavioral_random_percent_originali_condivisi}
    \end{subfigure}
    \caption{Comparison of content production and amplification dynamics for the \texttt{RandomRecommendation} configuration, illustrating \textit{(a)} the temporal evolution of first- and second-order actions and \textit{(b)} the cumulative percentage of original and re-shared content.}
    \label{fig:behavioral_random_contenuti_interazioni}
\end{figure}

Figure~\ref{fig:behavioral_random_percent_dirette_indirette_tempo} presents the percentage of first-order (purple) and second-order (yellow) actions across iterations for the \texttt{RandomRecommendation} configuration, while Figure \ref{fig:behavioral_random_percent_originali_condivisi} shows the temporal evolution of original versus re-shared content production.

\section{RQ3}
\subsection{Centrality Analysis} \label{app:centrality}

\begin{figure}[H]
     \centering
     \label{fig:rcp_degree_app}
     \subfloat[][]{\includegraphics[width=.51\columnwidth]{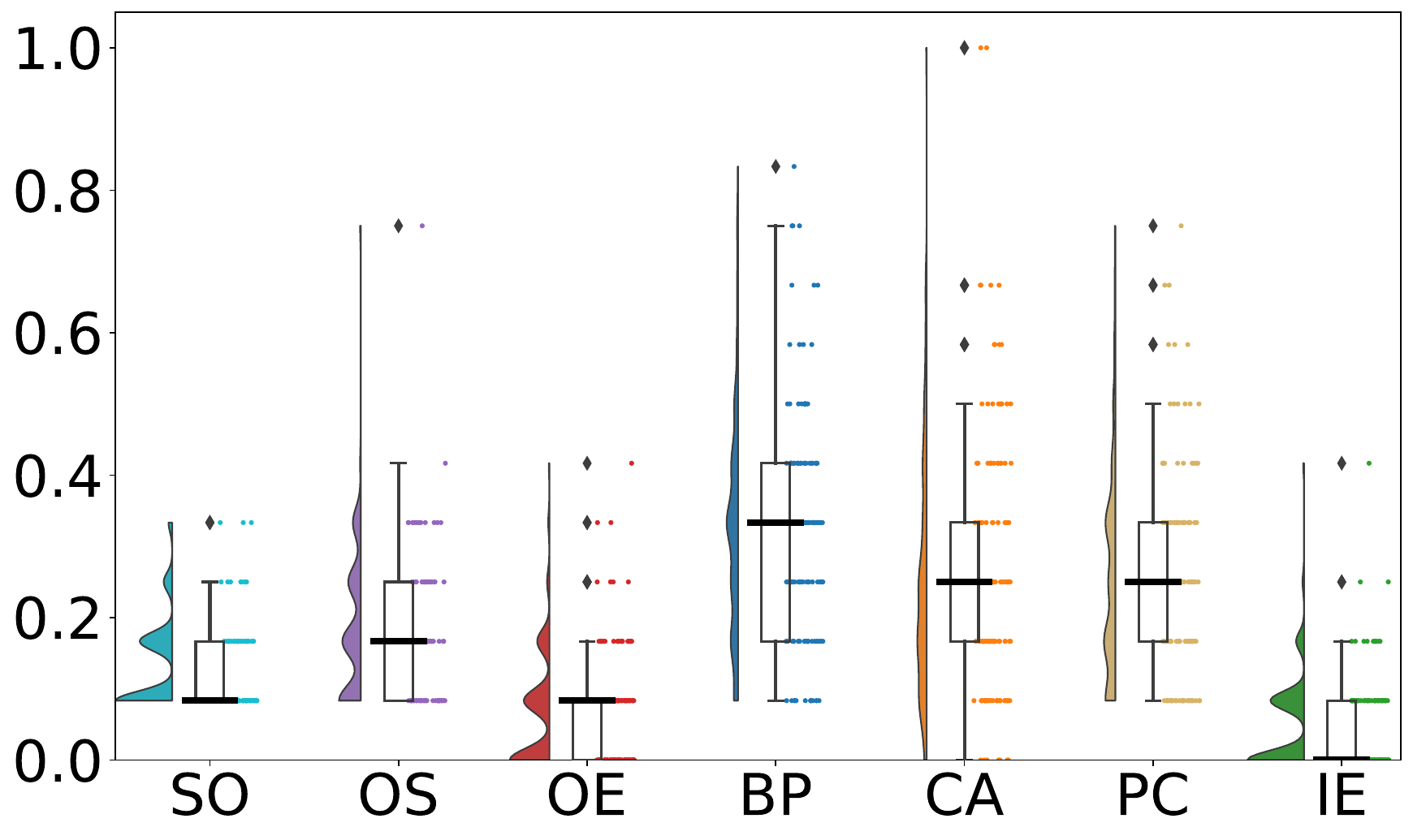}\label{fig:rcp_inDegree_interaction}}
     \subfloat[][]{\includegraphics[width=.51\columnwidth]{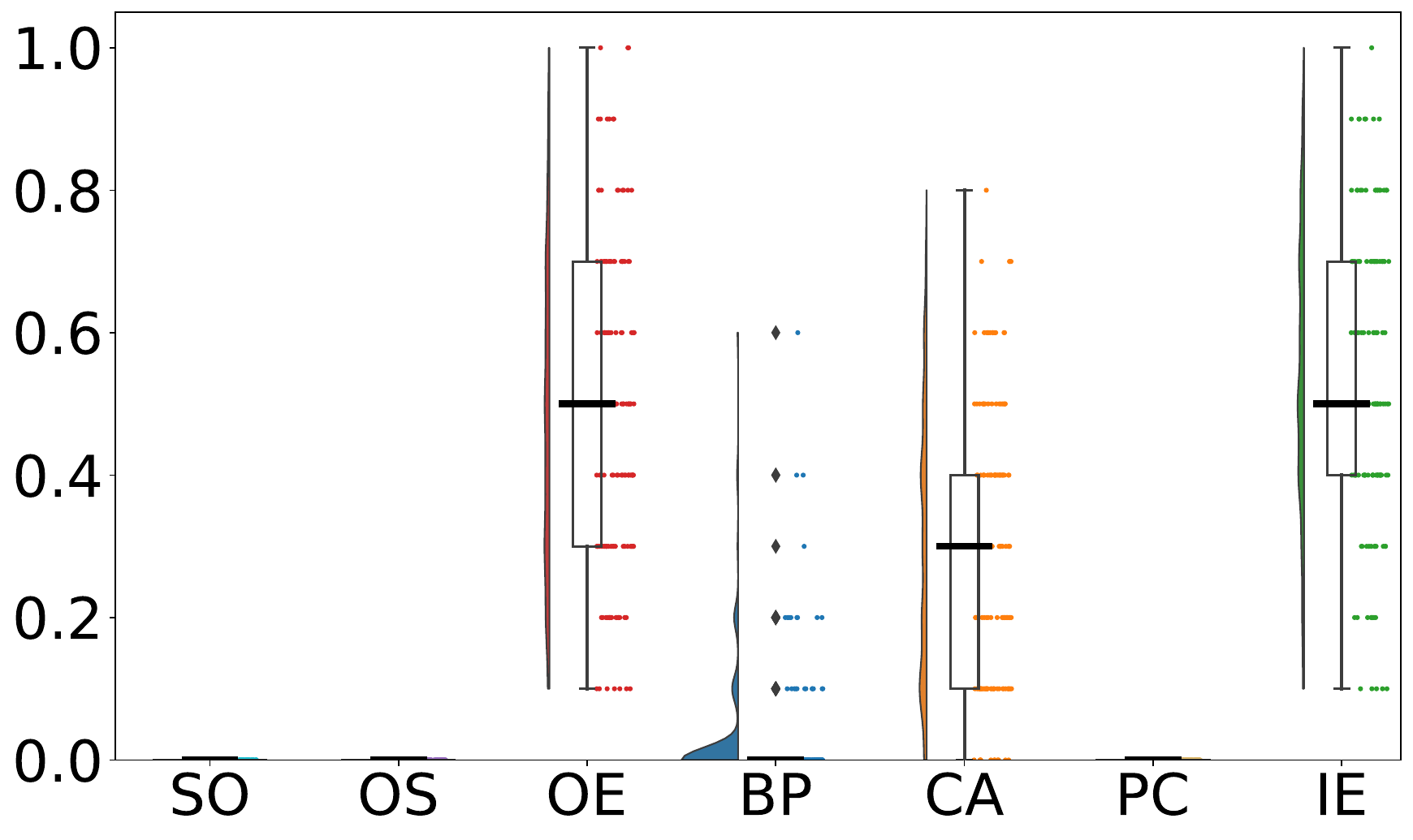}\label{fig:rcp_outDegree_interaction}}
    \caption{\textit{a)} In-degree and \textit{b)} out-degree centralities across behavioral traits in the interaction network.}
\end{figure}

Figures \ref{fig:rcp_inDegree_interaction} and \ref{fig:rcp_outDegree_interaction} show in-degree and out-degree centrality over the interaction network across behavioral traits for the \texttt{FullModel} configuration.

\section{Additional Results with Gemma 3} \label{app:gemma}

\subsection{Action Probability Across Behavioral Traits} \label{app:gemma_behav}

\begin{figure}[t]
    \centering
    \includegraphics[width=0.75\columnwidth]{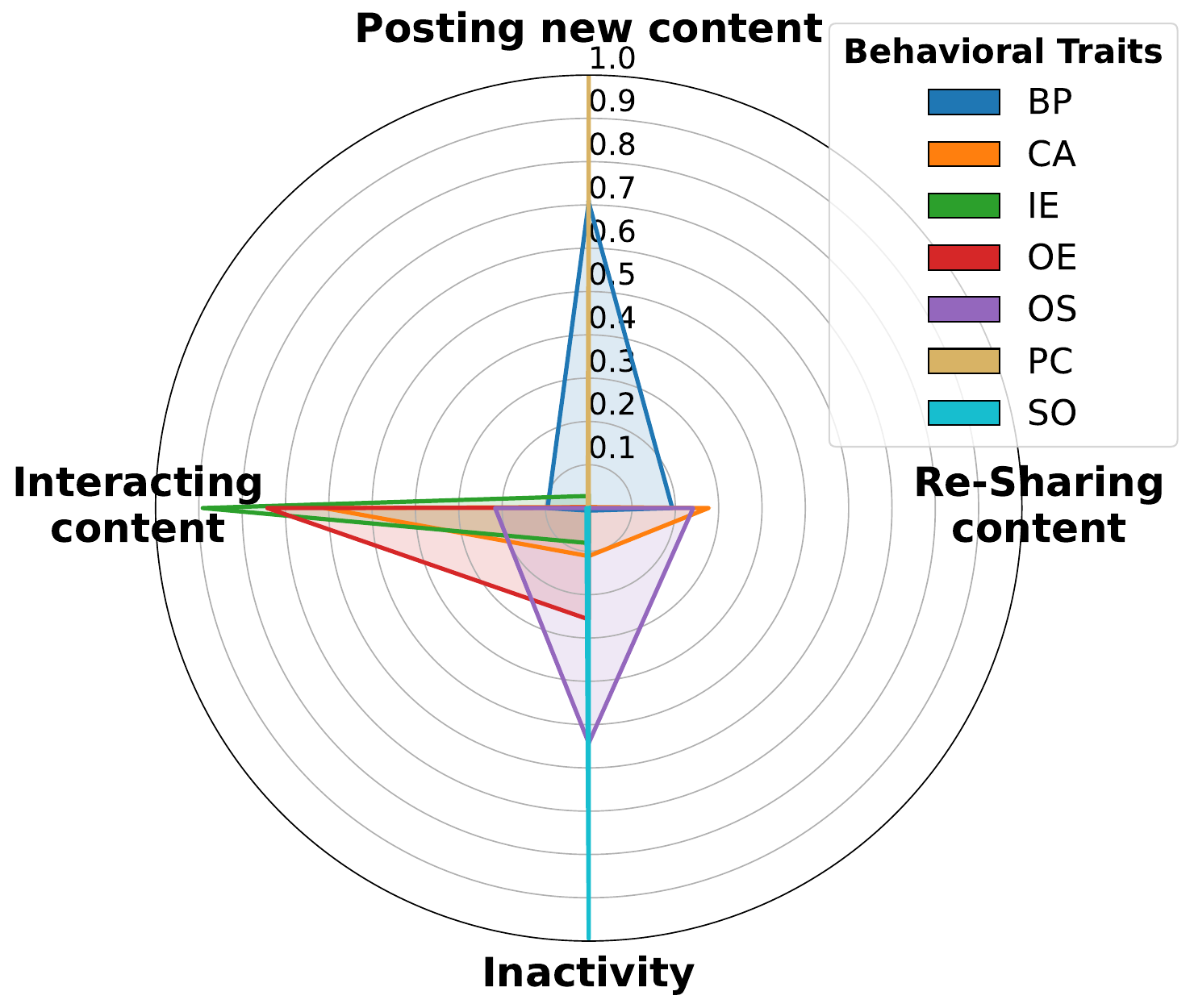}
    \caption{Average action probabilities across the seven behavioral traits for the simulation conducted with Gemma 3 27B.}
    \label{fig:radar_gemma}
\end{figure}

This subsection examines whether the behavioral patterns observed in RQ1 remain consistent when using a different underlying LLM. Figure~\ref{fig:radar_gemma} reports the average action probabilities across behavioral traits obtained from the simulation conducted with Gemma~3~27B. As shown in the figure, the resulting distributions closely mirror those observed in the main experiment, confirming that the emergence of profile-consistent behavioral patterns is driven by the behavioral trait design rather than by the specific language model employed. A slight difference can be observed in the relative balance between re-sharing and interaction actions: compared to the main experiment, agents tend to perform fewer re-sharing actions and a correspondingly higher number of interactions. This shift is most visible for amplification-oriented profiles, which, while maintaining their primary re-sharing role, exhibit a modest increase in interaction activity.

\subsection{Centrality Analysis} \label{app:gemma_centrality}
This subsection analyzes the centrality structure induced by behavioral traits when using Gemma~3~27B, focusing on both the re-sharing and the interaction networks.

\paragraph{Re-sharing Network.}

Figure~\ref{fig:rcp_degree} reports the distributions of normalized in-degree and out-degree centralities across behavioral profiles in the re-sharing network. Overall, the resulting patterns closely align with those observed in the main experiment. In particular, amplification-oriented profiles — Content Amplifiers (CA), Balanced Participants (BP), and Occasional Sharers (OS) — continue to occupy the most central positions in terms of out-degree, confirming their primary role in driving content propagation. A slight difference can be observed in the relative prominence of Content Amplifiers, whose out-degree centrality is marginally reduced compared to the main experiments. This effect is consistent with the behavioral patterns discussed in Appendix~\ref{app:gemma_behav}, where CA agents exhibit a modest reduction in re-sharing activity accompanied by increased interaction behavior. Despite this shift, the overall centrality ranking across behavioral profiles remains stable, indicating that the structural roles induced by behavioral traits are preserved.

\begin{figure}[t]
     \centering
     \subfloat[][]
     {\includegraphics[width=.24\textwidth]{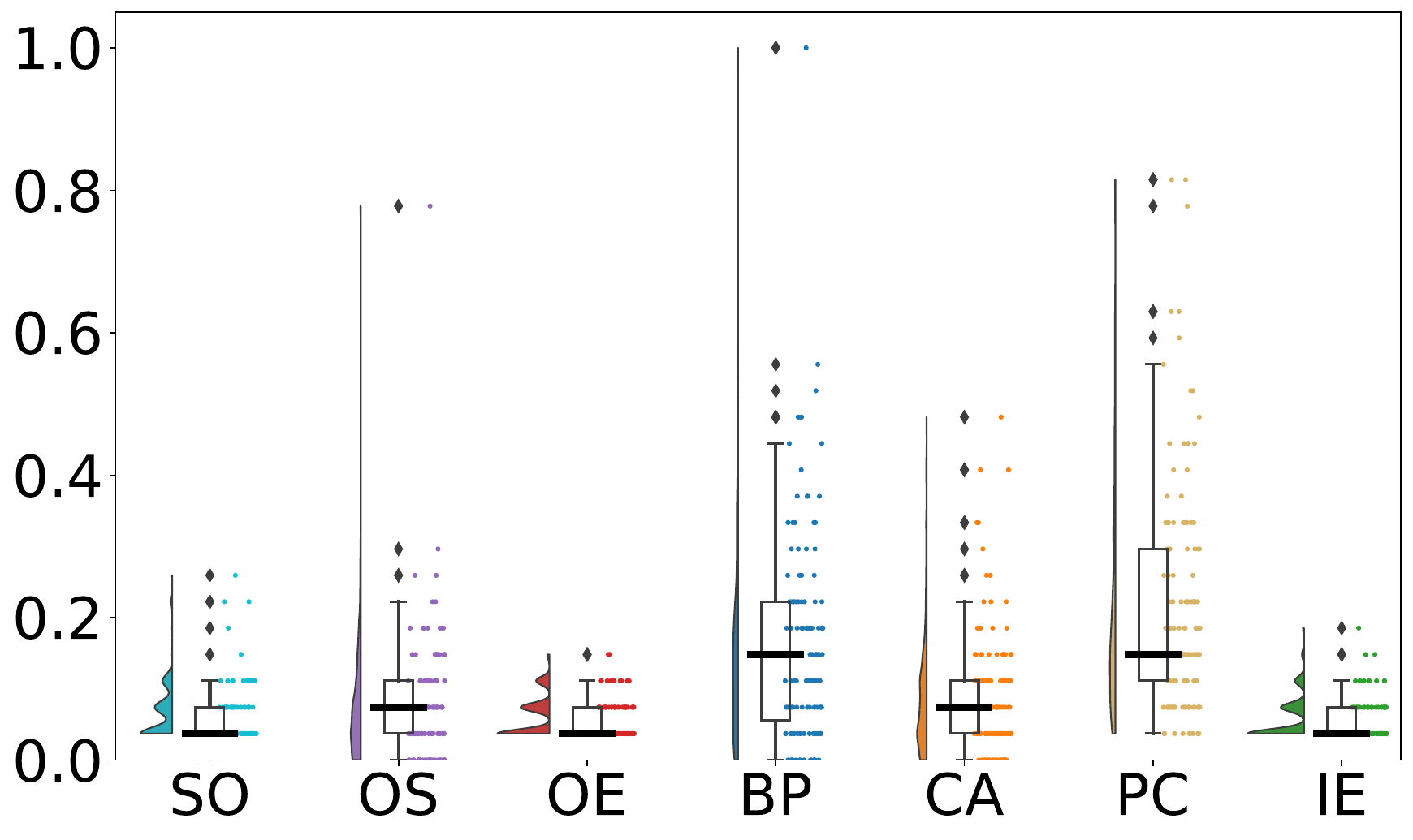}\label{fig:gemma_rcp_inDegree_retweet}}
     \subfloat[][]{\includegraphics[width=.24\textwidth]{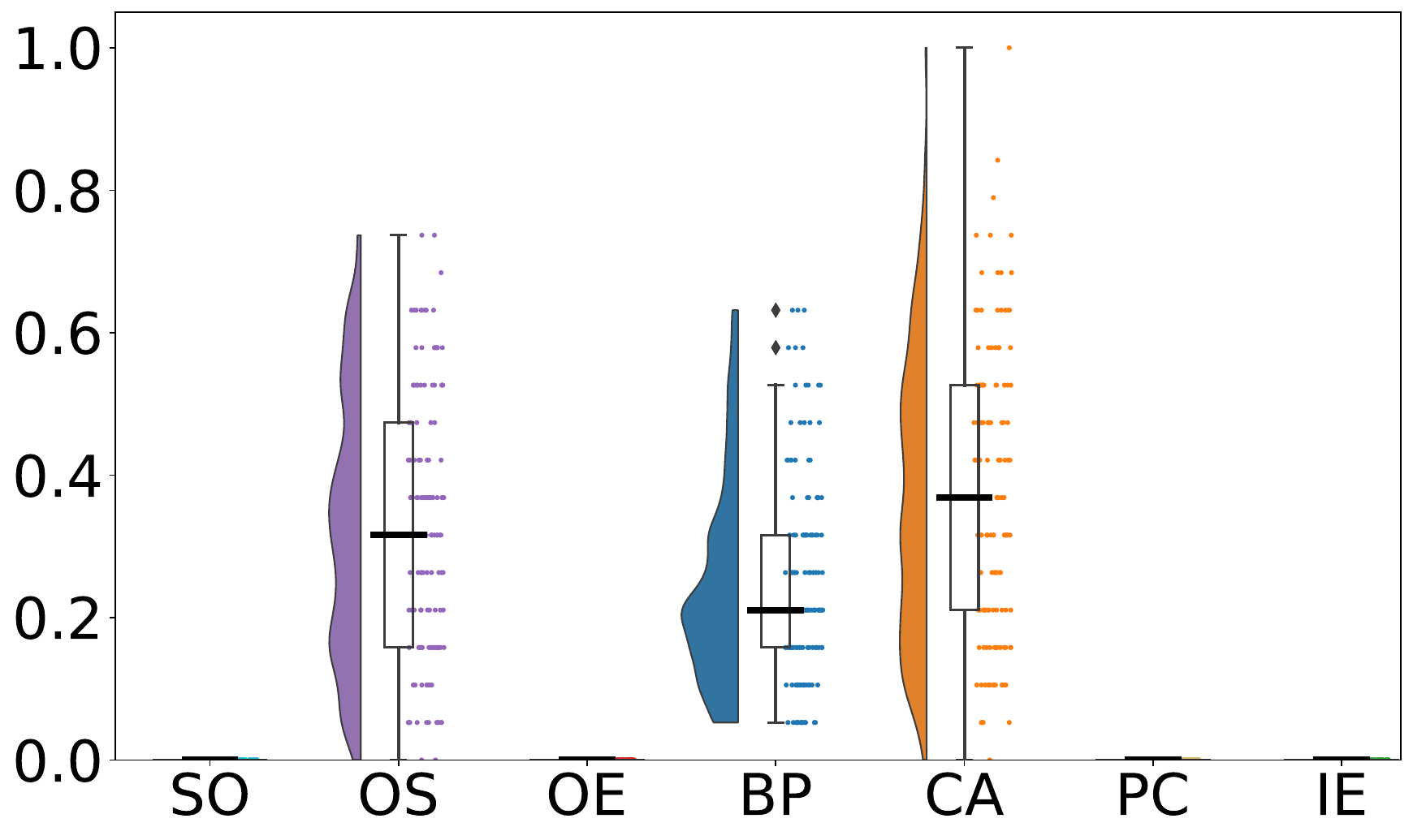}\label{fig:gemma_rcp_outDegree_retweet}}
     \caption{\textit{a)} In-degree and \textit{b)} out-degree centralities across behavioral traits in the re-sharing network for the simulation conducted with Gemma~3~27B.}   
     \label{fig:rcp_degree}
\end{figure}

\paragraph{Interaction Network.}

\begin{figure}[H]
     \centering
     \label{fig:rcp_degree_app}
     \subfloat[][]{\includegraphics[width=.24\textwidth]{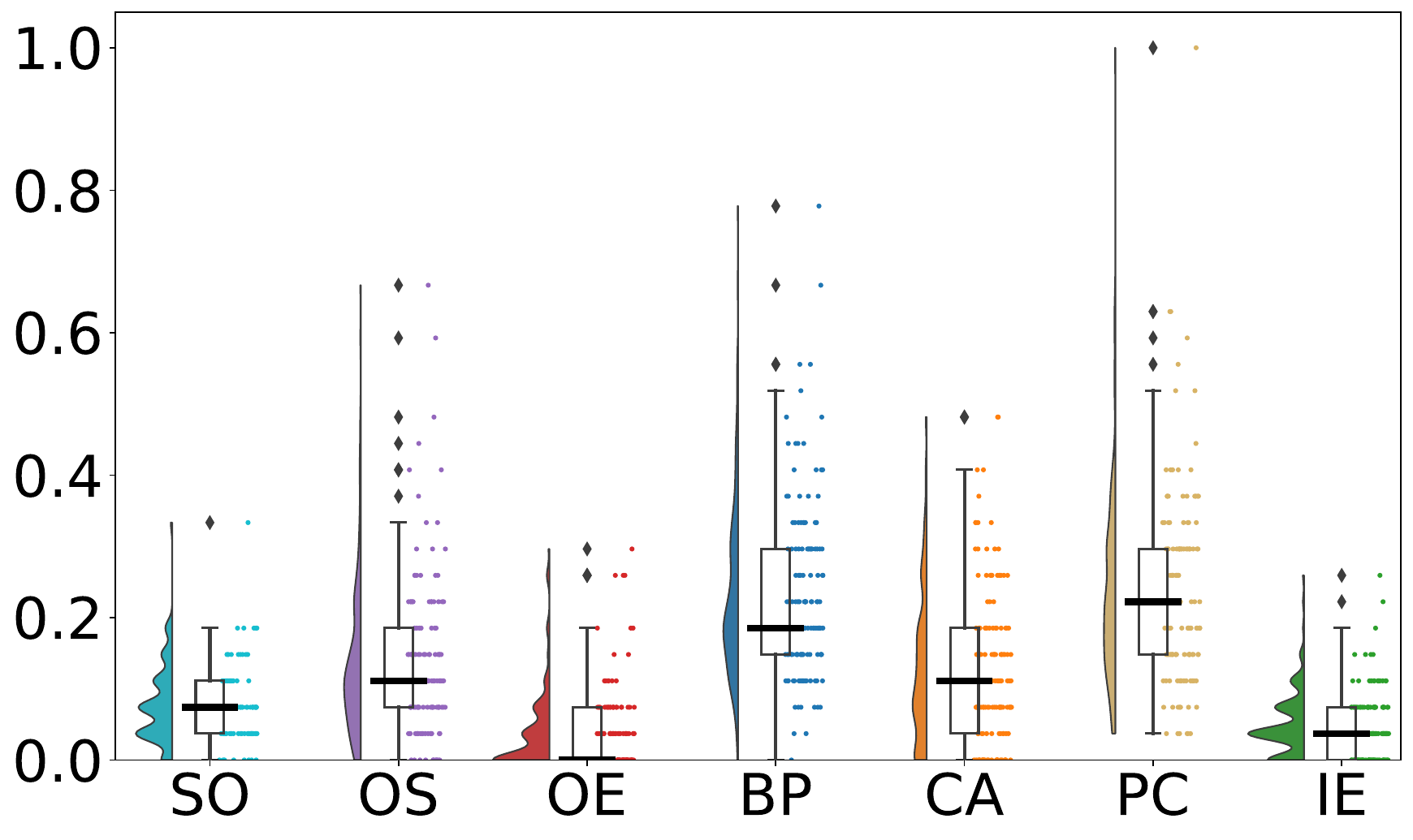}\label{fig:gemma_rcp_inDegree_interaction}}
     \subfloat[][]{\includegraphics[width=.24\textwidth]{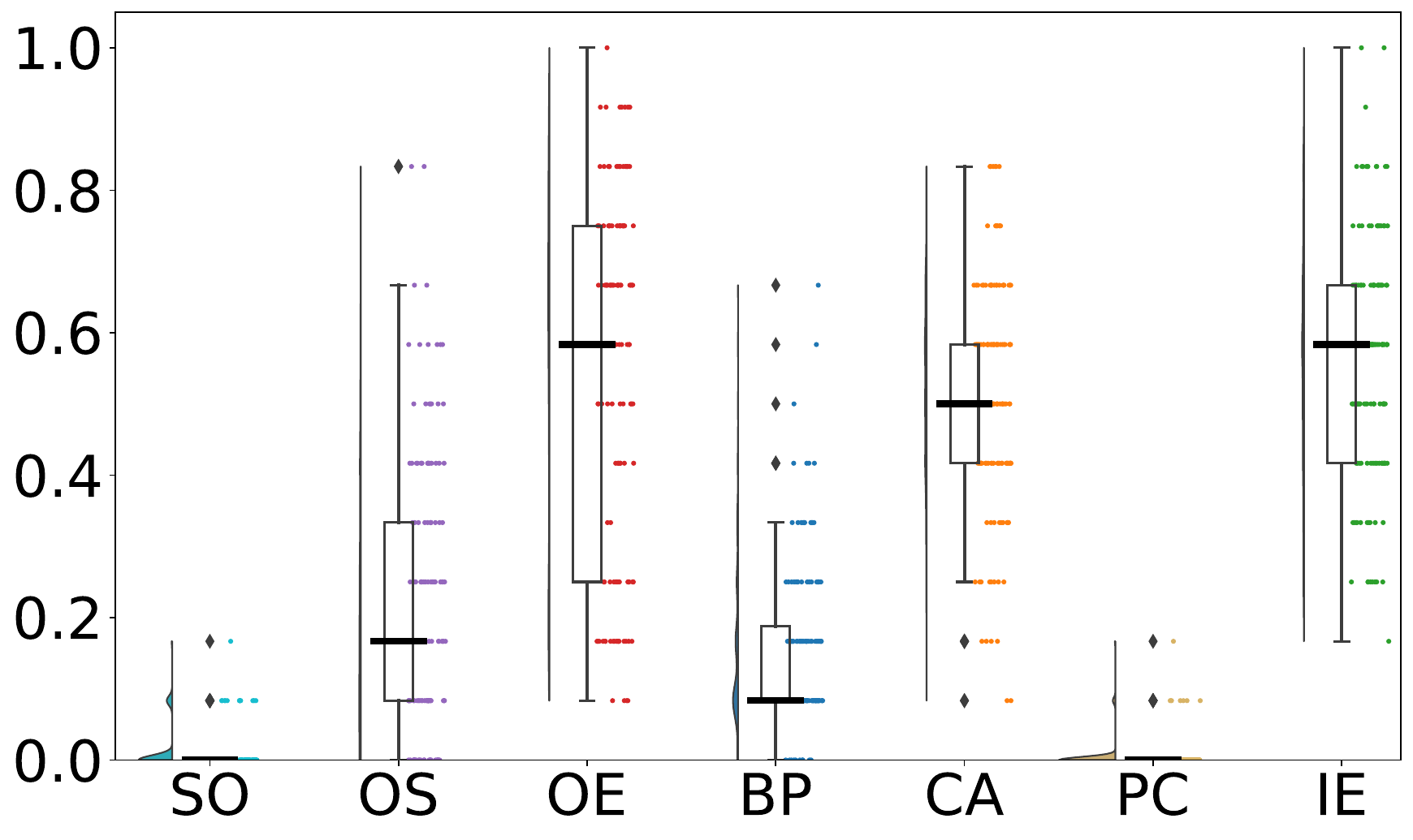}\label{fig:gemma_rcp_outDegree_interaction}}
    \caption{\textit{a)} In-degree and \textit{b)} out-degree centralities across behavioral traits in the interaction network for the simulation conducted with Gemma~3~27B.}
\end{figure}

The interaction network exhibits the degree centrality distributions shown in Figure~\ref{fig:gemma_rcp_outDegree_interaction}, computed across behavioral profiles. As in the main experiments, interaction-oriented profiles — Interactive Enthusiasts (IE) and Occasional Engagers (OE) — remain the most central in terms of out-degree, confirming their dominant role in generating interactions.

At the same time, minor differences emerge in the distribution of interaction activity across other profiles.
Occasional Sharers (OS), which exhibited near-zero interaction activity in the main experiments, show a higher level of out-degree centrality.
Balanced Participants (BP) and Content Amplifiers (CA) similarly display a modest increase in interaction activity, consistent with the reduced emphasis on re-sharing observed in this simulation.
Overall, however, the centrality structure of the interaction network remains aligned with that obtained using the main LLM, confirming the robustness of the observed patterns across different underlying language models.

\section{{\color{black}t-SNE Visualization of Empirical Action Probabilities}} \label{app:t_sne}

\begin{figure}[t]
    \centering
    \includegraphics[width=\columnwidth]{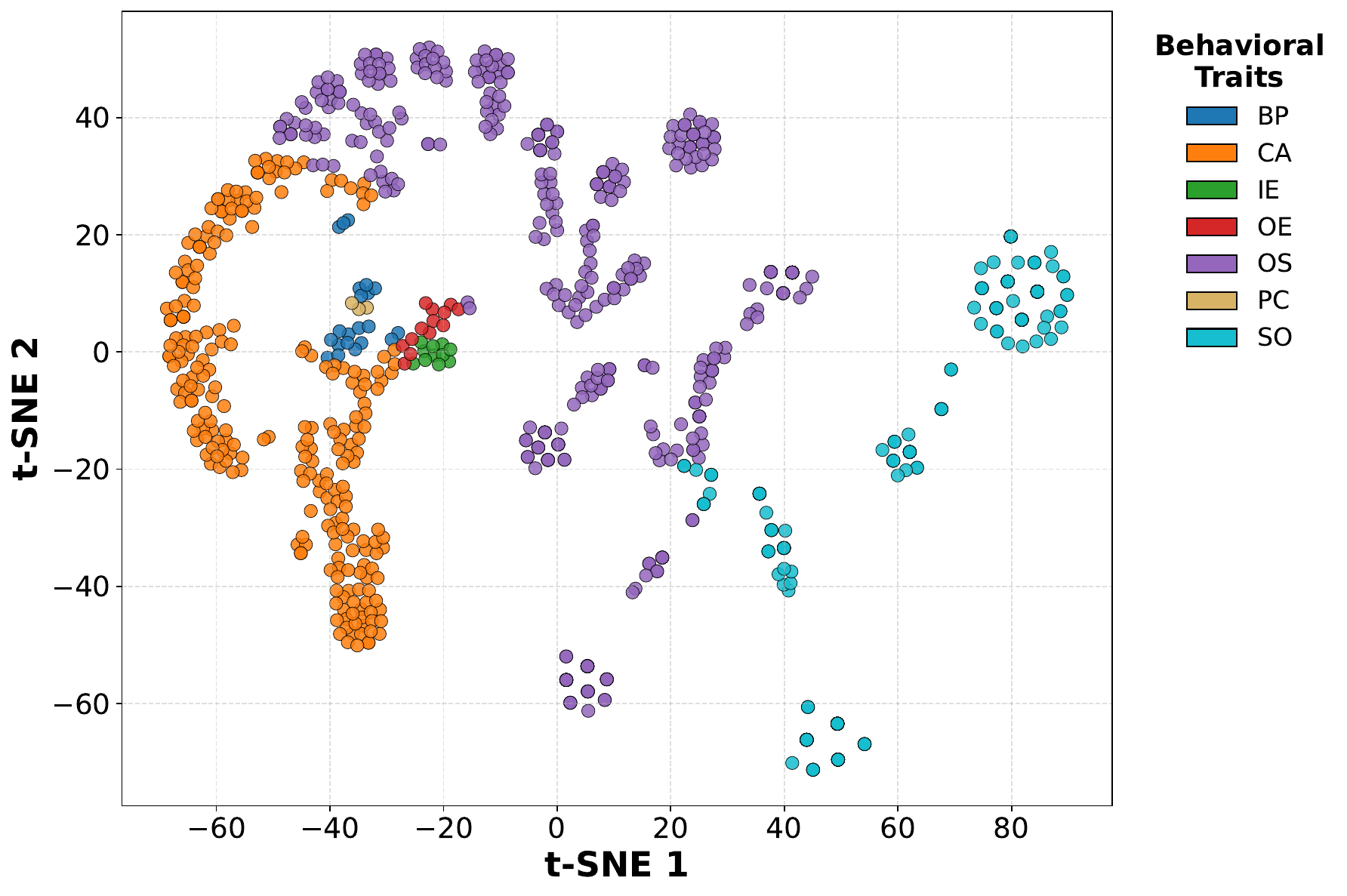}
    \caption{{\color{black}t-SNE projection of the four-dimensional empirical action-probability vectors of users. Each point represents a single user; colors denote the behavioral trait assigned.}}
    \label{fig:scatter_tsne}
\end{figure}

{\color{black}We provide a low-dimensional visualization of the empirical action-probability vectors used to assign real users to behavioral traits. Each user is represented by a four-dimensional vector encoding the relative frequency of content generation, re-sharing, interaction, and inactivity. We project these vectors onto two dimensions via t-SNE and color each point by the behavioral trait assigned through the k-means procedure described in Section \ref{sec:real_case}.

Figure \ref{fig:scatter_tsne} shows that users occupy well-separated regions of the projected space, with each trait concentrated in a distinct area. The two amplification-oriented profiles, Content Amplifiers (CA, orange) and Occasional Sharers (OS, purple), form the two largest groupings, in line with their dominant frequency in the empirical population. Silent Observers (SO, cyan) sit on the periphery as compact clusters, consistent with their predominantly inactive behavior. The smaller profiles (Balanced Participants, Occasional Engagers, Interactive Enthusiasts, and Proactive Contributors) form tighter groupings near the center, in proportion to their smaller population sizes. Although k = 7 was fixed a priori to match the seven behavioral traits of our framework, the projection shows that this granularity is also reflected in the data.}

\end{document}